%% file: RandomisedSorting_v1.tex
\documentclass{article}

\PassOptionsToPackage{%
	style	= alphabetic%
}{biblatex}
\input{sorting_preamble_biblio.tex}

\usepackage{geometry}
\input{sorting_preamble_main.tex}

\DeclareMathOperator{\perm}{perm}
\DeclareMathOperator{\inv}{inv}

\newcommand{\mstar}{\mathord\star}
\newcommand{\wt}{\widetilde}

\title{\sffamily%
	A Randomised Approach to Distributed Sorting
}

\author{\sffamily Sam Olesker-Taylor}
\date{\sffamily \today}
\date{}

\begin{document}

\maketitle

\vspace{-4ex}

\renewcommand{\abstractname}{\sffamily Abstract}

\begin{abstract}
\noindent
We introduce and analyse a new, extremely simple, randomised sorting algorithm:
\begin{itemize}[noitemsep]
	\item 
	choose a pair of indices $\set{i, j}$ according to some distribution $q$;
	
	\item 
	sort the elements in positions $i$ and $j$ of the array in ascending order.
\end{itemize}
We prove that taking $q_{\set{i,j}} \propto 1/\abs{j - i}$ leads to an order-$n (\log n)^2$ sorting time.

The sorter trivially parallelises in the \emph{asynchronous} setting, yielding a linear speed-up. We also exhibit a low-communication, \emph{synchronous} version with a linear speed-up.

We compare and contrast with other sorters, and discuss some of its benefits,
particularly its robustness and amenability to parallelisation and distributed computing.
\end{abstract}

\small
\begin{quote}
\begin{description}
	\item [Keywords:]
	sorting network,
	randomised algorithms,
	parallel \& distributed computing
	
	\item [MSC 2020 subject classifications:]
	68P10;
	60C05, 60J27,
	68W10, 68W14, 68W20, 68W40
\end{description}
\end{quote}
\normalsize




\blfootnote{%
	Sam Olesker-Taylor%
\quad
	\href{mailto:sam.olesker-taylor@warwick.ac.uk}{sam.olesker-taylor@warwick.ac.uk}%
\\
	Department of Statistics, University of Warwick, UK%
}

\vspace{-4ex}

\sffamily

\setcounter{tocdepth}{1}

\makeatletter
\renewcommand\tableofcontents{%
	\@starttoc{toc}%
}
\makeatother
\section*{Contents}
\vspace*{-2ex}

\tableofcontents

\unboldmath
\normalfont



\section{Introduction}

\subsection{Overview}

This paper investigates \emph{randomised sorting algorithms} based on weighted graphs.

Sorting algorithms lie and the core of computer science, underpinning a vast range of applications, from database management to scientific computing.
Classical sorting methods are less suitable in the era of `big data', even if they have optimal sequential complexity.
They need to be robust, scalable and amenable to both parallelisation and distribution.

This paper introduces a novel class of randomised sorting algorithms, leveraging graph-based probabilistic mechanisms to achieve efficient sorting with these properties.

The study of sorting networks provides a foundational perspective on parallel and distributed sorting, with origins tracing back to Batcher's bitonic sorting network \cite{B:bitonic}.
Sorters such as bitonic \cite{B:bitonic} or parallel MergeSort \cite{C:parallel-mergesort} require $\Oh{\log n}$ parallel steps for merging, leading to a total of $\Oh{(\log n)^2}$ steps.
A theoretical breakthrough came when \textcite{AKS:sorting:conf,AKS:sorting:jour} gave a depth-$\Th{\log n}$ sorting network---itself a randomised construction.
Practically, $n \ge 2^{75}$ is needed for AKS to beat bitonic, though.

Deterministic sorting networks like bitonic and odd--even MergeSort have been widely implemented in hardware due to their predictable structure.
However, they require precise coordination across nodes, and are susceptible to worst-case inputs, limiting their utility in larger-scale or adversarial environments.

Randomisation can be added to address some of these challenges. A prominent example is randomised QuickSort: it achieves $\Th{n \log n}$ complexity for all inputs, avoiding the worst-case $\Th{n^2}$ for deterministic QuickSort.
Another example is random sampling in MergeSort.

The introduction of \emph{randomised} sorting networks significantly broadened the scope of parallel sorting techniques.
\textcite{AHRV:random-sorting-networks} analysed the probabilistic behaviour of uniformly random sorting networks.
More recently, work has focused on extending these principles to distributed and fault-tolerating settings---settings in which our particular proposal excels.
\textcite{GSS:prob-bubble-sort} achieved robust sorting in dynamic, multi-agent systems with the use of randomised algorithms by introducing introduced self-replicating population protocols.
\textcite{AFHN:adversarial-sorting:jour,AFHN:adversarial-sorting:conf} address sorting with adversarial comparisons; in particular, they show a lower bound of $\Om{n^{1+\eps}}$ for deterministic algorithms.
\textcite{T:sorting-selection-adversary} builds on this work, deriving a randomised sorting algorithm needing $\Oh{n (\log n)^2}$ comparisons, thus outperforming deterministic counterparts.

Randomisation has been key in addressing scalability in distributed settings.
Eg, parallel MergeSort \cite{C:parallel-mergesort} uses random sampling to optimise performance in shared-memory architectures,
and randomised load-balancing techniques have been applied \cite{AHS:counting-networks}.

Building on this body of work, the algorithm we present in this paper selects a pair $\set{i,j}$ (with $i \ne j$) according to some distribution $q$ and sorts the entries in positions $i$ and $j$.
The key is to choose $q$, and then runtime $T$, smartly:
we simply require
\[
	1/q_{i,j} = \Oh{ \abs{i - j} }
\Quad{and}
	T = \Thb{n (\log n)^2}.
\]
We also provide two parallel implementations, yielding a linear speed-up for up to $n/4$ cores.

The approach has several key benefits, which we outline very briefly here.

\begin{description}
	\item 
	[Simplicity.]
	The algorithm is extremely easy to implement, including in distributed settings.
	
	\item 
	[Robustness and Fault Tolerance.]
	Randomisation makes it naturally resilient to adversarial inputs.
	It is also highly tolerant to faults, unlikely highly structured approaches, in which errors propagate through recursive stages.
	
	\item 
	[Scalability and Parellisation.]
	Each comparison operates independently.
	This makes it highly compatible with parallel and distributed systems,
	whether asynchronous or not.
	
	\item 
	[Load Balancing.]
	Each element in the array is inspected roughly the same number of times.
\end{description}

\noindent%
A more detailed comparison of sorting algorithms is given in \S\ref{sec:intro:comparison}.

\subsection{Set-Up and Main Results}

For clarity, when we talk about the complexity of a \emph{randomised} algorithm, we mean that the bound holds both in expectation and with polynomially high probability in the length~$n$.

We start by defining the framework of \textit{graph-based randomised sorters}.

\begin{defn}
\label{def:intro:graph}
	%
Let $V$ be a finite set with an implicit total ordering.
Let $G = (V, w)$ be a weighted graph; here, $w : \set{ \set{i,j} \given i,j \in V, i \ne j } \to \coi{0, \infty}$.
We write $E \cq \set{ e \mid w(e) \ne 0 }$ for the set of (non-null) edges, and $w(i,j)$ for $w(\set{i,j})$ when $i,j \in V$ with $i \ne j$.

A single step of the \textit{randomised sorter} chooses an edge $e \in E$ according to $w$%
	---ie, edge $e \in E$ is chosen with probability $w(e) / \sumt{e' \in E} w(e')$---%
and sorts the cards at its endpoints.

More precisely, suppose that $e = \set{i, j}$ with $i < j$ is chosen in a given step:
\begin{itemize}[noitemsep]
	\item 
	swap the cards if the label of the card in position $i$ is larger than that in position $j$;
	
	\item 
	otherwise, do nothing, leaving the cards in ascending order.
\end{itemize}
	%
\end{defn}

\begin{rmkt}
\label{rmk:intro:robust}
This graph-based randomised sorter is robust to mistakes.
Suppose that the sorting mechanism may be faulty:
	it correctly sorts pair $(i,j)$ with probability $p_{i,j} \ge p$;
	otherwise, it does nothing---it never `unsorts' a pair.
All our proofs apply to this model, and the same bounds hold except with the runtimes multiplied by $1/p$.
\end{rmkt}

We start by analysing the most basic procedure: choose a pair uniformly at random.

\begin{thm}
\label{res:intro:unif}
Let $n \in \mbn$.
A single step of the \textit{uniform sorter} chooses $i,j \in [n]$ uniformly at random without replacement and sorts the cards at $\set{i,j}$.

The uniform sorter
requires order $n^2 \log n$ comparisons to sort a length-$n$ list in the worst case,
both in expectation and with polynomially small-failure probability.
\end{thm}

This corresponds to the complete graph,
which is one extreme.
The other is the path.

\begin{thm}[{\cite[Theorem~12]{GSS:prob-bubble-sort}}]
Let $n \in \mbn$.
A single step of the \textit{adjacent sorter} chooses $k \in [n-1]$ uniformly at random and sorts the cards at $\set{k, k+1}$.

The adjacent sorter
requires order $n^2$ comparisons to sort a length-$n$ list in the worst case,
both in expectation and with polynomially small-failure probability.
\end{thm}

Both the complete graph and path are very slow,
but for different reasons.

\begin{description}
	\item [Path.]
	Each edge is chosen at rate $1/(n-1) \approx 1/n$; this is as fast as possible since the graph must be connected, so $\abs E \ge n-1$.
	However, a single card is only ever moved distance $1$, and thus may need to be touched $n-1 \approx n$ times to reach its target~position.
	
	\item [Complete Graph.]
	On the other hand, in the complete graph, regardless of a card's current position, there is an edge connecting it to its target position.
	However, getting closer to the target does not help: the required edges is always chosen at rate $1/\binom n2 = \Th{1/n^2}$.
\end{description}

\noindent%
In short, the complete graph has long-range sorting, whilst the path only does short-range.

We desire a graph that handles \emph{all} scales simultaneously to get $n \poly\log n$ comparisons.

\begin{thm}
\label{res:intro:harm}
	%
Let $n \in \mbn$.
The \textit{harmonic sorter} is the randomised sorter
with
\[
	V
\cq
	[n]
\Qand
	w(\set{i,j})
\cq
	1/\abs{j - i}
\Qfor
	i,j \in [n]
\text{ with }
	i \ne j.
\]

The harmonic sorter
requires order $n (\log n)^2$ comparisons to sort a length-$n$ list in the worst case,
both in expectation and with polynomially small failure probability.
\end{thm}

\begin{rmkt}
\label{rmk:intro:harm}
Detailed intuition for this construction is given in \S\ref{sec:outline}.
In short,
\[
	\log 2
\le
	w\rbb{ 0, \coi{2^{k-1}, 2^k} }
\le
	1
\Qforall
	k \in \set{1, ..., \floor{\lg n}},
\Qwhere
	\lg n \cq \log_2 n.
\]
So, if the scales grow exponentially, it is equally fast to compare $0$ with any scale.
Roughly, this allows the sorter to \textit{approximately halve}:
	get most of the cards with label $\ell < n/2$ to a position $i < n/2$ and those with label $k \ge n/2$ to a position $j \ge n/2$, in order $w(E) = \Th{n \log n}$ comparisons.
If the halving were \textit{perfect}, then we would simply iterate $\lg n$ times.
It is not perfect, but this gives some intuition for the order $n (\log n)^2$ sorting~time.
\end{rmkt}


Our last results concern \emph{synchronised} parallel sorters, in which a disjoint collection of pairs is sorted in each step. They are, in essence, parallelisations of the harmonic sorter.

\begin{defn}
\label{def:intro:par}
A \textit{matching} in $[n]$ is a collection of disjoint pairs in $[n]$.
Let $\tilde q$ be a probability measure over the set of matchings in $[n]$.
A step of the \textit{parallel randomised sorter} chooses a matching $M \sim \tilde q$ and sorts the cards at the endpoints of each edge (ie, pair) $e \in M$.
\end{defn}

\begin{figure}[t]
\centering
\includegraphics[width = \linewidth]{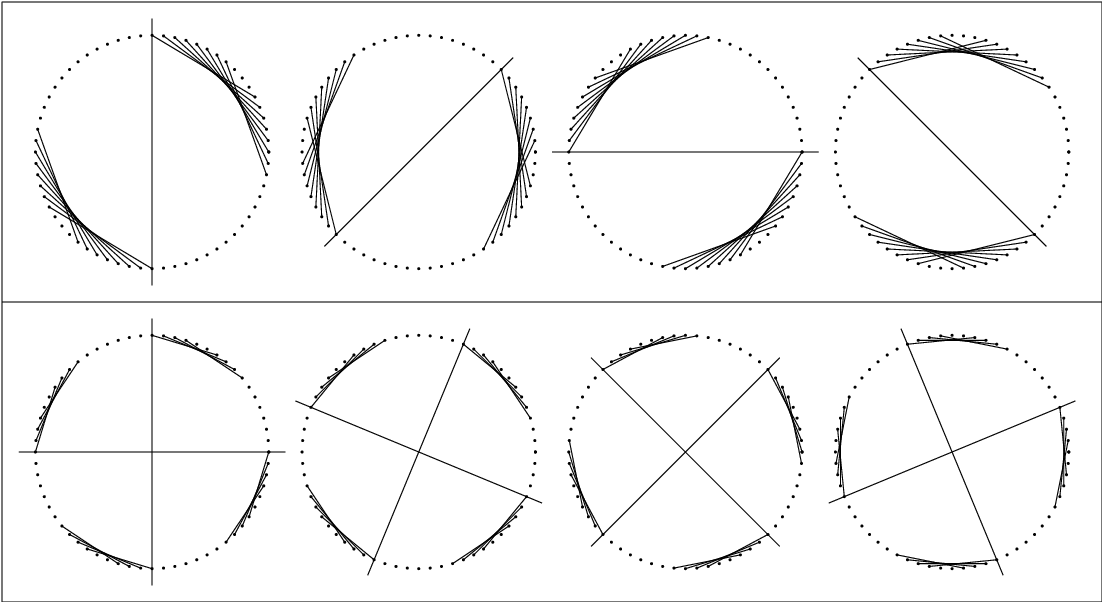}
\caption[Two possible matchings, with $n = 64$]{%
	Two examples of collections $(M_0, M_1, M_2, M_3)$ of matchings, both with $n = 64$.
	\begin{itemize}
		\item \textsf{Above.}
		Edge-lengths $D = 11$, so $K = \floor{\lg(n/D)} = 2$.
		Each of the four matchings has $2^{K-1} = 2$ `blocks', each containing $n/2^{K+1} = 8$ edges of length $D = 11$.
		\item \textsf{Below.}
		Edge-lengths $D = 6$, so $K = \floor{\lg(n/D)} = 3$.
		Each of the four matchings has $2^{K-1} = 4$ `blocks', each containing $n/2^{K+1} = 4$ edges of length $D = 6$.
	\end{itemize}
	The blocks are all rotations of the `fundamental block' $E_{D,K}$; the bisecting straight lines indicate the starting point of this rotated block.
	Each matching contains $n/4 = 16$ edges
	}
\label{fig:intro:matchings}
	%
\end{figure}

We analyse two parallelisations.
	The first is deterministic, for $p = n/4$ cores; it is visualised in \cref{fig:intro:matchings}.
	The second is random, for any $p \le n/4$ cores.
We use the notation
\[
\textstyle
	E_{d,k}
\cq
	\bigcup_{i=0}^{n/2^{k+1}-1}
	\set{i, i + d}
\text{ for }
	k, d \in \mbn
\Qand
	x + S
\cq
	\set{ x + s \given s \in S }
\text{ for }
	x \in \mbr
\text{ and }
	S \subseteq \mbr.
\]
Here, and below, addition is calculated mod $n$, as are differences and distances.


\begin{thm}
\label{res:intro:par:n}
	%
Let $N \in \mbn$ and set $n \cq 2^N$.
Define $\tilde q$ by sampling matching $M$ as follows.

\begin{enumerate}
	\item 
	Sample $K \sim \Unif([\lg n])$ and, conditionally, sample $D \sim \Unif( \oci{n/2^{K+1}, n/2^K} )$.
	
	\item 
	Define $M_0 \cq \bigcup_{\ell=0}^{2^{K-1}-1} (\ell n/2^{K-1} + E_{D,K})$ and $M_r \cq r n/2^{K+1} + M_0$ for $r \in \set{1, 2, 3}$.
	
	\item 
	Sample $R \sim \Unif(\set{0, 1, 2, 3})$ and set $M \cq M_R$.
\end{enumerate}
Then,
$\abs M = \tfrac14 n$ deterministically.

The randomised sorter associated to $\tilde q$
requires order $(\log n)^2$ rounds to sort a length-$n$ list in the worst case,
both in expectation and with polynomially small failure probability.
\end{thm}


\begin{thm}
\label{res:intro:par:p}
Let $n, p \in \mbn$ with $p \le \tfrac14 n$.
Define $\tilde q$ by sampling matching $M$ as follows.

\begin{enumerate}
	\item 
	Sample $E_1, ..., E_p \sim^\iid w$ where $w(\set{i,j}) = 1/\abs{j - i}$.
	
	\item 
	Set $M \cq \bigcup_{k=1}^p \set{ E_k \given E_k \cap E_\ell = \emptyset \: \forall \, \ell \ne k }$.
\end{enumerate}
Then,
$\abs M \le p$ deterministically and $\tfrac12 p \le \ex{\abs M} \le p$.

The randomised sorter associated to $\tilde q$
requires order $(n/p) (\log n)^2$ rounds to sort~a~length-$n$ list in the worst case,
both in expectation and with polynomially small failure~probability.
\end{thm}


We now explain how to implement these algorithms in a computational environment.


\begin{rmkt}
\label{rmk:intro:par:n}
Suppose there are $p = n/4$ computational nodes.
Sampling the matching in \cref{res:intro:par:n} requires a centralised controller to sample $K$, $D$ and $R$, then send these to the computational nodes.
This sampling requires order $\log n$ bits.
With just this information, and its own label, each node can easily determine which comparison it must undertake.
\end{rmkt}

\begin{rmkt}
\label{rmk:intro:par:p}
Suppose there are $p \le n/4$ computational nodes.
For \cref{res:intro:par:p},
each node samples an edge according $w$---ie, $\set{i,j}$ is chosen with probability proportional to $1/\abs{j - i}$---which requires order $\log n$ bits.
The actual update takes place in two rounds.

\begin{enumerate}
	\item 
	In the first round, each node places a mark on each endpoint of the edge it has selected.
	
	\item 
	After this has been done, each node checks if was the only node to mark either endpoint of its edge.
	If this is the case, then it sorts that pair; otherwise, it does nothing.
\end{enumerate}
The key point is that, conditional on one node choosing $\set{i,j}$,
the chance that either $i$ or $j$ are chosen by other nodes is at most $2p/n \le \tfrac12$, by the union bound and marginal uniformity.

In fact, our proof is more flexible than this, provided \textit{atomic sorting steps} are used:
	when $\set{i,j}$ is sorted, both locations $i$ and $j$ are locked until the sort action is complete;
	any other computational node requesting the value at these locations must wait until the sort is complete.
This way, all the $p$ pairs can be sorted---even those intersecting each other.
This (stochastically) reduces the number of rounds required, but atomic actions are slower.
\end{rmkt}




We close with a couple of open questions.
As explained in \S\ref{sec:outline},
the harmonic sorter~is~inspired by the hypercube $\mbz_2^N$, with vertices relabelled according to its Gray code.
The connectivity is `rigid': eg, $0$ connects to the last vertex in each sub-dimension, to $1$, $3$, $7$, ...,~$2^N - 1$.

The structural rigidity of the hypercube is not amenable to our proof.
The intuition for the harmonic sorter was to relax this rigidity, spreading the weight evenly across each sub-dimension, rather than connecting only to a specific one.
This is still straightforward to sample from, and yields a simple parallelisation. But, it would be nice to be able to analyse the hypercube sorter too.
It parallelises particularly easily, too:
simply use dimension-cuts.


\begin{openq*}
	%
Let $N \in \mbn$. Let $V \cq \mbz_2^N$, with edges $E$ defined by the Gray code.
Let
\[
	M_k
\cq
	\cup_{v \in \mbz_2^N}
	\set{ x, x + e_k }
\Qwhere
	(e_k)_\ell
\cq
	\one{ k = \ell }
\text{ for }
	\ell \in [N]
\Qfor
	k \in [N].
\]
\begin{enumerate}
	\item 
	Let $w(e) \cq 1$ for $e \in E$.
	The probability $\set{i,j}$ is chosen is $2/(n \lg n)$, where $n \cq 2^N$.

	\item 
	Sample $M \sim \tilde q$ by drawing $K \sim \Unif([N])$ and letting $M \cq M_K$.
	Always, $\abs M = \tfrac12 n$.
\end{enumerate}

The randomised sorters from \cref{def:intro:graph,def:intro:par} associated to $w$ and $\tilde q$
require order $n (\lg n)^2$ comparisons and $(\log n)^2$ rounds, respectively, to sort a length-$n$ list in the worst case,
both in expectation and with polynomially small failure probability.
\end{openq*}

The other major open question is, ``Can we do better than $n (\log n)^2$ in this framework?'' A natural target is $n \log n$.
It is our (soft) belief that $n (\log n)^2$ is the fastest---and hence \cref{res:intro:harm} is optimal.
In fact, we feel that the hypercube is actually the ideal construction.

The reason is that if edges are required between all scales, then each vertex has (weighted) degree $\log n$, and seeing a given comparison between scales takes order $n \log n$ comparisons. There is likely then a coupon-collector argument which imposes a further factor of $\log n$.

\begin{openq*}
There exists a constant $c > 0$ such that,
for all $n \in \mbn$ and any weighted graph $G = ([n], w)$,
there exists a length-$n$ list
that the associated randomised sorter requires at least $c n (\log n)^2$ comparisons to sort,
with probability tending to $1$ as~$n \to \infty$.
\end{openq*}

\subsection{Comparison of Sorting Algorithms}
\label{sec:intro:comparison}

There is more literature on sorting than can be repeated here, or even in a dedicated survey.
We focus on benefits of randomisation in general, then on those particular to \emph{our} algorithm.



\subsubsection*{Randomised vs Deterministic}

We highlight several advantages of randomised sorters.

\begin{description}
\item 
[Robustness Against Adversaries.]
Randomised algorithms ensure robustness by relying on unpredictable choices which an adversary cannot anticipate or exploit.
%


\item 
[Resource Optimisation.]
Randomness is are often amenable to distributed computing set-ups,
and can help balance load across computation nodes in such cases; eg,
	choosing pivot/division points in QuickSort/MergeSort
or
	randomising the gap in ShellSort.

\item 
[Practical Performance.]
Randomised algorithms often outperform deterministic ones in real-world scenarios due to their average performance typically being sufficient.
Eg, randomised QuickSort typically outperforms deterministic MergeSort \cite{S:algorithm-design-manual}.
\end{description}

\subsubsection*{Benefits of Our Randomised Sorting Algorithm}

See \cref{tab:intro:sorting-comparison} for a summary.

\begin{description}
\item 
[Simplicity of Implementation.]
The sorting procedure is extremely simple.
No recursive structure, or maintaining of auxiliary data structures (eg, heaps or merge trees), is needed.

\item 
[No Synchronisation.]
Many algorithms, such as QuickSort or MergeSort, and particularly bitonic sort,
require explicit synchronisation or coordination;
eg, ``Sort these subarrays, then merge.''
Contrastingly, every step of our algorithm has the same description.

\item 
[Distributed-System Compatibility.]
Structured algorithms often require frequent exchange of data, possibly with auxiliary data structures, which is expensive in a distributed system; eg, merging subarrays.
In our algorithm, nodes operate almost independently.

\item 
[Natural Load Balancing.]
If the weighted graph is close to regular, then every element is inspected roughly the same number of times. So, the load is automatically balanced.

\item 
[Fault Tolerance.]
The algorithm does not need to know if a given sorting step was successful or not, only the probability of being successful; see \cref{rmk:intro:robust}.
Contrast this with MergeSort or bitonic, in which errors in sorting subarrays propagate.
\end{description}

\begin{table}[t]
\centering
\renewcommand{\arraystretch}{1.3} 
\begin{tabular}{|p{1.55cm}|p{2.95cm}|p{3cm}|p{4.7cm}|}
	\hline
	\textbf{\textsf{Algorithm}}
&	\textbf{\textsf{Data Exchange}}
&	\textbf{\textsf{Synchronisation}}
&	\textbf{\textsf{Our Algorithm's Advantage}}
\\ \hline
	\textsl{\textsf{Bitonic}}
&	High \newline (frequent merging)
&	High \newline (strictly staged)
&	Minimal communication and asynchronous operation
\\ \hline
	\textsl{\textsf{Bubble}}
&	Low \newline (local comparisons)
&	Low \newline (sequential passes)
&	Non-local comparisons provide linear speed up
\\ \hline
	\textsl{\textsf{Heap}}
&	Moderate \newline (heap operations)
&	Moderate \newline (heap maintenance)
&	Avoids heap complexity, fully independent operations
\\ \hline
	\textsl{\textsf{Quick/ Merge}}
&	Moderate \newline (partitioning)
&	Moderate \newline (recursive calls)
&	No partitioning or recursion, better for distributed use
\\ \hline
	\textsl{\textsf{Radix/ Counting}}
&	High \newline (bucketising data)
&	Moderate \newline (global buckets)
&	General-purpose, avoids reliance on specific key structure
\\ \hline
\end{tabular}
\caption{Summarised comparison of sorting algorithms}
\label{tab:intro:sorting-comparison}
	%
\end{table}

\subsection{Notation}

We use the following standard mathematical notation.
\begin{itemize}
	\item 
	Let $f, g : \mbn \to \mbr_+$.
	Then, 
		$f = \Oh g$ if $\limsup_{n\to\infty} f(n)/g(n) < \infty$.
	and
		$f = \Th g$, or $f \asymp g$, if $0 < \liminf_{n\to\infty} f(n)/g(n) \le \limsup_{n\to\infty} f(n)/g(n) < \infty$.
	
	\item 
	Let $X$ and $Y$ be random variables taking values in $\mbr_+$.
	Then, $X \lesssim Y$ if there exists a coupling such that $\pr{X \le Y} = 1$, or equivalently if $\pr{ X \ge t } \le \pr{ Y \ge t }$ for all $t \ge 0$.
\end{itemize}

\section{Outline}
\label{sec:outline}

We outline the key steps in the proof.
First, we reduce to a simpler problem; see \S\ref{sec:prelim:reduction}.

\begin{Proof}[Reduction]
\qedtriangle
The 0--1 principle, adjusted to randomised sorting networks, can be used to reduce to handling only 0--1-valued sequences.
By padding the start/end of the sequence with 0-s/1-s, respectively, we may assume that the 0--1 sequence has length a power of $2$ and is \textit{balanced}---ie, has the same number of 0-s as 1-s.
We assume this from now on.
Sorting such a sequence requires moving all the 0-s to the back half and the 0-s to the front half.
\end{Proof}

We now turn to the task of moving the 0-s and 1-s to their respective target halves.

\begin{Proof}[Uniform Sorter]
\qedtriangle
For the uniform sorter, we do this directly, in essence in a single argument.
For every 0 in the back half---ie, with position $i \ge n/2$---there is 1 in the front half---ie, with position $j < n/2$.
Let $M_t$ denote the number of these \textit{misplaced} 0-s (or 1-s) at time $t$.

For every misplaced 0, it needs to be swapped with a misplaced 1.
Hence, there are precisely $M_t^2$ pairs that need swapping.
So, in a single step, $M_t \to M_t - 1$ with probability $M_t^2 / \binom n2$, remaining unchanged otherwise.
The sorting time is precisely $\inf\set{ t \ge 0 \given M_t = 0 }$.

By the summability of $\sum_{m\ge1} 1/m^2$, this has expectation order $n^2$.
However, we need a polynomially small failure probability, which requires an extra $\log n$ factor.
\end{Proof}

The harmonic sorter is significantly more challenging to handle.
First, we give intuition.

\begin{Proof}[Intuition]
\qedtriangle
We have seen that both the path and complete graphs give rise to slow sorting. However, this is for fundamentally different reasons:
	the complete graph performs long-range, approximate sorting well,
	but its short-range sorting is slow;
	the path performs short-range sorting well,
	but has no long-range edges.
We desire a graph which sorts well at all scales.

A candidate is the hypercube $G = \mbz_2^N$, with vertices enumerated by Gray codes, rather than the usual binary (ie, lexicographic) ordering.
Briefly, the $N$-th Gray code $\mathsf{Gray}_N$ is defined by concatenating two copies of the ($N-1$)-th Gray code:
	the first is prepended with a $0$
and
	the second with a $1$, and also reversed.
Concretely,
	$\mathsf{Gray}_0$ is the empty sequence $\emptyset$,
	$\mathsf{Gray}_1 = (0, 1)$,
	$\mathsf{Gray}_2 = (00, 01, 11, 10)$
and
	$\mathsf{Gray}_3 = (000, 001, 011, 010, 110, 111, 1001, 100)$.

In particular, the first vertex $0...0$ connects to the \emph{last} vertex of every sub-dimension: eg, for $N = 4$, the 0-th vertex $0 = 0000$ connects to $1 = 0001$, $3 = 0010$, $7 = 0100$ and $15 = 1000$.
The long-range edges allow the sorter to get the card into approximately the right location, before the short-range edges take over to do the final local corrections.

Each vertex has degree $N = \lg n$, so the total number of edges is $\tfrac12 n \lg n$. In the unweighted case (ie, $w(e) = 1$ for all edges $e$), this means each edge is picked at rate $2/(n \log n)$.

Altogether, this makes it a good candidate for $n \poly n$ sorting.
However, the rigid structure of the hypercube can make it hard to work with: a vertex in the first half (of the form $0\star$) connects to a \emph{specific} vertex in the second half.
The harmonic sorter is a relaxation of this, in some sense:
	vertex $0...0$ connects to \emph{all} vertices in the second half,
	spreading its weight roughly equally amongst all $n/2$,
	and similarly for the other sub-cubes.
\end{Proof}

Next, we explain how to use this mixture of long- and short-range edges to sort.
It is convenient to rescale the labels and positions by $n$, so that both lie in $\coi{0,1}$ rather than $[n]$.

\begin{Proof}[Using Long- and Short-Range Edges Iteratively]
\qedtriangle
The strategy is not to move a misplaced 0---ie, one in $\coi{\tfrac12, 1}$---\emph{directly} into $\coi{0, \tfrac12}$.
Indeed, the long-range edges have low weight, so this swap may have probability order $1/(n^2 \log n)$, eg if the current state is $10...01...10$.

Instead, the misplaced 0-s/1-s are iteratively moved closer and closer to $\coi{0, \tfrac12}$/$\coi{\tfrac12, 1}$, respectively, \emph{roughly} having their distance away in each iteration.
Eg, for $10...01...10$, it is easy to move the misplaced 0 (at the end) \emph{close} to the front half.
As the misplaced labels get closer to their target half, the number of labels they can swap with gets smaller, but this is balanced (perfectly) by the increasing of the weight of the shortening edges.
\end{Proof}

Next, we describe a slightly naive approach, yielding an order $n (\log n)^3$ bound;
see~\S\ref{sec:harm:1}.

\begin{Proof}[Naive: Order $n (\log n)^3$]
\qedtriangle
We first `clean' the \emph{entirety} of $\coi{0, \tfrac14}$/$\coi{\tfrac34, 1}$ of 1-s/0-s, respectively.
Once $\coi{0, \tfrac14}$/$\coi{\tfrac34, 1}$ consists of no 1-s/0-s, respectively, no comparison involving $\coi{0, \tfrac14} \cup \coi{\tfrac34, 1}$ can ever cause a swap.
So, it remains to sort the `inner' half of the sequence.
Sorting this is done analogously:
	first, clean $\coi{\tfrac14, \tfrac38}$ and $\coi{\tfrac58, \tfrac34}$,
	then the inner quarter remains.
This must be iterated $\lg n$ times, after which point the full sequence is sorted.
The `scale-invariance' highlighted in \cref{rmk:intro:harm} means each iteration takes roughly the same length of time.

In a single step, a given misplaced 0 is moved out of $\coi{\tfrac34, 1}$ with probability order $1/(n \log n)$.
This \emph{looks like} it leads to an order $n \log n$ `cleaning' time for $\coi{0, \tfrac14} \cup \coi{\tfrac34, 1}$; this would lead to an order $n (\log n)^2$ sorting time.
However, a coupon-collector argument means an extra $\log n$ factor is needed to \emph{completely} clean $\coi{0, \tfrac14} \cup \coi{\tfrac34, 1}$.
\end{Proof}

Finally, we refine this argument to get the true order $n (\log n)^2$ bound;
see~\S\ref{sec:harm:2}.

\begin{Proof}[Refined: Order $n (\log n)^2$]
\qedtriangle
The previous approach was an iterative halver.
However, do we really need the \emph{entirety} of $\coi{0, \tfrac14} \cup \coi{\tfrac34, 1}$ to be clean before we start cleaning $\coi{\tfrac14, \tfrac38} \cup \coi{\tfrac58, \tfrac34}$?

The answer is clearly ``no''.
Instead, we  wait until $\coi{0, \tfrac14} \cup \coi{\tfrac34, 1}$ is 99\% clean---which does not require an extra $\log n$ factor---before starting on $\coi{\tfrac14, \tfrac38} \cup \coi{\tfrac58, \tfrac34}$.
The key challenge now is to argue that this remaining `dirty' 1\%
does not overly impact the future cleaning.
Arguing this carefully is technically challenging.
This gives the desired order $n (\log n)^2$ bound.
\end{Proof}

It remains to discuss the parallel implementations.

\begin{Proof}[Parallel Implementations]
\qedtriangle
Given a collection of \emph{vertex-disjoint} edges (ie, a matching) to be applied, the order in which the edge-sortings are applied is irrelevant.
So, a (parallel) matching-based implementation can still be viewed as an ordered sequence of comparisons.

The difficulty in viewing such `parallel' Markov chains in this `sequential' manner is that the `sequential chain' is no longer Markovian: the choice of edge in each step is not independent.
Nevertheless, our proof is highly robust to these changes.
All that really matters is that the probability that pair $\set{i,j}$ is chosen as part of a matching behaves like $1/\abs{j - i}$.

The reason for this is the way the technical aspect of the `sequential' proof works.
Very roughly, there arise certain sums of random variables which are independent in the sequential case, but not in the parallel framework.
However, the proof only relies on their expectation (and tail via Markov's inequality).
So, the lack of independence does not matter.
\end{Proof}

We make two final comments on the nature of the proofs, before proceeding formally.

\begin{Proof}[Continuous-Time]
\qedtriangle
The `sequential' proofs use the analogous \emph{continuous-time} process:
\begin{itemize}[noitemsep]
	\item 
	compare, and sort, the endpoints of each edge $e \in E$ at rate $w(e)$;
	\item 
	
	in particular, \emph{some} comparison is undertaken at rate $w(E)$.
\end{itemize}
It is straightforward to convert a continuous-time result into an analogous discrete-time one,
using Poisson concentration,
since the claimed sorting times are of the form $\Th{ w(E) \log n }$.

For clarity, we exclusively use \textit{time} to refer to units of continuous time; eg, ``in time $T$'' means $T$ units of continuous time, in which (roughly) $T w(E)$ comparisons are made.

Using continuous time makes the robustness in \cref{rmk:intro:robust} clear:
	proposing $\set{i,j}$ at rate $w(\set{i,j})$ but only accepting with probability $p_{i,j}$
is \emph{identical} to
	proposing at rate $p_{i,j} w(\set{i,j})$ and always accepting.
Moreover, our proofs only rely on \emph{lower} bounds on the rates.
\end{Proof}

\begin{Proof}[Space Rescaling]
\qedtriangle
For simplicity, we rescale the labels and locations by a factor $n$,
\[
	\text{from}
\quad
	[n] = \set{0, ..., n-1}
\Quad{to}
	\coi{0,1}_n \cq \set{0, 1/n, ..., 1 - 1/n} \subseteq \coi{0,1}.
\qedhere
\]
	%
\end{Proof}

A reader who is comfortable with the set-up may safely skip to the harmonic sorter in~\S\ref{sec:harm}:
\begin{itemize}[noitemsep]
	\item 
	the only result used from \S\ref{sec:prelim:reduction} is that only balanced, 0--1 strings need be considered;
	
	\item 
	the analysis of the uniform sorter in \S\ref{sec:unif} is not referenced, serving more as a warm-up.
\end{itemize}
This completes the outline.
The remainder of the paper consists of the formal proofs.

\section{Reduction}
\label{sec:prelim:reduction}

The purpose of this section is to prove the following simplification.
Once we get to the proofs of the main theorems (in \S\ref{sec:unif} and \S\ref{sec:harm}), we assume this simplification, without comment.

\begin{reduction}
\label{res:prelim:reduction}
To establish the claimed bounds on the random sorting times,
we may assume the following,
provided the failure probability established is at most $1/n^2$.
\begin{itemize}
	\item 
	The sequences have length which is a power of $2$.
	
	\item 
	The sequences are 0--1 valued, with half the entries 0 and half 1.
\end{itemize}
\end{reduction}

The 0--1 principle
states that a sorting network correctly sorts all lists of a given length if and only if it sorts all such 0--1-valued lists.
Our sorters are \emph{probabilistic}, though: there is no grand coupling such that \emph{all} length-$n$ lists are sorted by \emph{the same} sequence of comparisons.
The 0--1 principle extends to this set-up, at the cost of a factor $n$ to the failure probability.

\begin{prop}
Let $n \in \mbn$ and $T \in \mbn$.
Let $\mcn \cq \rbr{ \set{i_t, j_t} }_{t=1}^T \in \binom{[n]}2{}^T$ be an ordered collection of pairs and $x \in \mbr^n$ a sequence.
For $k \in \set{0, ..., n}$, define $x_{(k)} \in \set{0,1}^n$ by replacing the $k$ largest entries of $x$ by $1$, breaking ties arbitrarily, and the remainder by $0$.
\begin{itemize}
	\item 
	$\mcn$ sorts $x$ if and only if it sorts $x_{(k)}$ for all $k$.
	
	\item 
	$\mcn$ sorts $x$ whp if it sorts $x_{(k)}$ with probability $1 - \oh{1/n}$, for each $k$.
\end{itemize}
\end{prop}

\begin{Proof}
Each $x_{(k)}$ is a projection of $x$, so sorting $x$ clearly sorts $x_{(k)}$ for all $k$.

The converse is a simple inductive argument.
We prove it in the case that all entries of $x$ are distinct, but the same argument works in general---just with a little more housekeeping.

Let $x'$ and $x'_{(k)}$ denote the sequences after application of $\mcn$.
If $x_{(1)}$ is sorted, then the last element of $x'$ must be the largest.
Similarly, if $x_{(2)}$ is sorted, then the last two elements of $x'$ contain the largest two elements of $x$; but, the largest element is last, so the penultimate element of $x'$ must be the second largest element of $x$.
Iterating, the $k$-th largest element is in the $k$-th to last position.
This is true for all $k$, so $x'$ must in fact be sorted.

Regarding the probabilistic aspect, if $x'_{(k)}$ is sorted with probability $1 - \oh{1/n}$, for each $k$, then all $n+1$ are simultaneously sorted with probability $1 - \oh{1}$, by the union bound.
\end{Proof}

We now need only sort 0--1 sequences.
Moreover, by padding the beginning/end with 0-s/1-s, we may assume that half of the entries are 0 and half 1, at the cost of increasing $n$.

\begin{defn}
Given even $m, k \in \mbn$, define $\set{0,1}^m_k \cq \set{ y \in \set{0,1}^m \given \sum_i y_i = k }$.

Let $n \in \mbn \setminus \set{0}$ and $k \in \mbn$;
let $n_0 \cq 2^{\ceil{\lg n}} - n + k$, $n_1 \cq 2^{\ceil{\lg n}} + k$ and $\wt n \cq n + n_0 + n_1 = 2^{\ceil{\lg n} + 1}$.
Define the \textit{lift} $\lambda : \set{0,1}^n_k \to \set{0,1}^{\wt n}_{\wt n/2}$ by
	prepending $n_0$ 0-s to the start
and
	appending $n_1$ 1-s to the end,
and translating the indices:
\[
	\lambda(x)_i
\cq
\begin{cases}
	x_i
&\text{for}\quad
	i \in V \cq \set{1, ..., n}
\\
	0
&\text{for}\quad
	i \in V_0 \cq \set{-(n_0-1), ..., 0},
\\
	1
&\text{for}\quad
	i \in V_1 \cq \set{n+1, ..., n+n_1}.
\end{cases}
\]
The weighted graph $\wt G = (V \cup V_0 \cup V_1, \wt w)$ is a \textit{lift} of $G = (V, w)$ if $\fnrestrict{\wt w}{E} = w$.
\end{defn}

All our bounds are at most polynomial in $n$.
So, replacing $n$ by $\wt n \le 4n$ is no loss.

\begin{lem}
Let $n \in \mbn \setminus \set{0}$.
Let $x \in \set{0,1}^n$, and let $k \cq \sum_{i \in [n]} x_i$ denote the its number of 1-s.
Set $\wt x \cq \lambda(x)$, and let $\wt G$ be any lift of $G$.
Let $(X_t)_{t\ge0}$ and $(\wt X_t)_{t\ge0}$ denote the processes driven by $G$ and $\wt G$, started from $x$ and $\wt x$ respectively.
Then, there exists a coupling such~that
\[
	\lambda(X_t)
=
	\wt X_t
\Qforall
	t \ge 0.
\]
In particular,
$X_t$ is sorted if and only if $\wt X_t$ is also sorted.
\end{lem}

\begin{Proof}
Crucial to this is the fact that the 0-s/1-s in $V_0$/$V_1$ can never move:
\[
	\wt X_t(i)
=
\begin{cases}
	0
&\text{for all}\quad
	i \in V_0
\\
	1
&\text{for all}\quad
	i \in V_1
\end{cases}
\Qforall
	t \ge 0.
\]
So, we can use the trivial coupling, remembering that $\fnrestrict{\wt w}{E} = w$:
\begin{itemize}
	\item 
	whenever a pair $(i, j)$, with $i,j \in [n]$, is compared in $G$,
	also compare that pair in $\wt G$;
	
	\item 
	sample the remaining comparisons independently.
\end{itemize}
These extra comparisons have one endpoint in $V_0 \cup V_1$, and hence do not affect the state.
\end{Proof}

We prove the theorems by showing that any 0--1 length-$n$ sequence is sorted in the claimed time with probability $1 - \Oh{1/n^2}$.
By the union bound, an arbitrary length-$n$ sequence is sorted in the claimed time (and hence order of comparisons) with probability $1 - \Oh{1/n}$.

\begin{itemize}
	\item 
	We could easily adjust the proof to obtain an arbitrarily high power in the exponent, say $1 - \Oh{1/N^C}$.
	The cost would be a linear multiple in the exponent $C$.
	For simplicity of presentation, we prove it for a fixed $C$ only.
	
	\item 
	Since $1 - \Oh{1/n} \ge \tfrac12$, the sorting time (of a given length-$n$ sequence) can be upper bounded by a Geometric random variable with mean $2$, multiplied by the claimed time.
	This Geometric has expectation $2$, and so the expected sorting time is of the same order at the whp-sorting time.
	So, it suffices to bound the whp-sorting time only.
\end{itemize}

For the remainder of the paper, we assume that the sequence
	is 0--1-valued,
	has length $n$ of the form $n = 2^N$, for some $N \in \mbn$,
and
	has the same number of 0-s as 1-s.

\section{Uniform Sorter}
\label{sec:unif}

In this section, we analyse the \textit{uniform sorter} which chooses a pair of distinct locations $i$ and $j$ uniformly at random and relatively sorts the cards at these positions.
This corresponds to letting $G$ be the unweighted (ie, $w(e) = 1$ for all $e \in E$) complete graph in \cref{def:intro:graph}.
In particular, note that $w(E) = \binom n2 \asymp n^2$.
So, $n^2 \log n$ \emph{comparisons} takes order $\log n$ \emph{time}.

This section should be seen as a warm-up for the main event (namely, the harmonic sorter), helping the reader get familiar with the type of continuous-time arguments used.
It is safe for a confident reader to skip the section entirely, if desired.

\begin{thm}[\cref{res:intro:unif}]
\label{res:unif:sorting-time}
The uniform sorter
requires order $n^2 \log n$ comparisons to sort any length-$n$ list,
both in expectation and with polynomially high probability.
\end{thm}

The lower bound is short and simple, so we give it straight away.

\begin{Proof}[Proof of \cref{res:unif:sorting-time}: Lower Bound]
Start from permutation $x = (2, 1, 4, 3, 6, 5, ...)$.
Then, the only pairs $(i,j)$ which have any effect are of the form $(2k-1, 2k)$. Once each such pair has been applied (at least) once, the deck is sorted.
Each is selected at rate $1$, so it takes time order $\log n$ to select all $\floor{n/2}$ such pairs, by the standard coupon-collector argument.
	%
\end{Proof}

We use continuous time to prove the upper bound:
	each pair $(i,j) \in [n]^2$ with $i < j$ is selected at rate $1$ (simultaneously).
In this scaling, we must show an order-$\log n$ sorting~time.
Recall also that we are dealing only with 0--1 sequences with an equal number of 0-s and~1-s.

\begin{defn}
For $t \ge 0$,
let $M_t$ be the number of misplaced 0-s (or 1-s):
\[
	M_t
\cq
	\absb{ X_t^0 \setminus [n/2] }
=
	\absb{ X_t^1 \setminus ([n] \setminus [n/2]) }
\]
where $X_t^{0/1}$ is the set of locations of the $0$-s/$1$-s at time $t$, respectively.
\end{defn}

Observe that $t \mapsto M_t$ is (deterministically) non-increasing under the dynamics: a 0 can only move to the left (or stay in place), so cannot move out of its target set $[n/2]$.
The strategy behind the proof is always to bound the rate at which the number $M_t$ decreases.
The sorting time, which we denote $\tau_\star$, is precisely the first time at which $M_t = 0$.

%

\begin{prop}
\label{res:unif:rate-hit}
Let $E_m \sim \Exp(m^2)$ independently for $m \ge 1$.
Then, $\tau_\star$ satisfies
\[
	\tau_\star
\lesssim
	T_\star
\cq
	\sumt{m\ge1}
	E_m.
\]
\end{prop}

\begin{Proof}
Rather than drawing \emph{locations} $(i, j)$, with $i < j$, we equivalently draw \emph{labels} $(k, \ell)$, with $k < \ell$.
We assume that their positions satisfy $X_t(k) > X_t(\ell)$,
else no swap happens.
\begin{itemize}
	\item 
	If $k \ge \tfrac12$ or $X_t(\ell) \ge \tfrac12$,
	then the count is unaffected.
	\begin{itemize}
		\item 
		If $\ell \ge \tfrac12$,
		then both cards have labels not in $\coi{0, \tfrac12}$.
		
		\item 
		If $X_t(\ell) \ge \tfrac12$,
		then both cards have positions not in $\coi{0, \tfrac12}$.
	\end{itemize}
	
	\item 
	Else,
	if $\ell \le \tfrac12$ or $X_t(k) < \tfrac12$,
	then the count is still unaffected.
	\begin{itemize}
		\item 
		If $X_t(k) < \tfrac12$,
		then both cards labelled $k$ and $\ell$ remain in $\coi{0, \tfrac12}$.
		
		\item 
		If $\ell \le \tfrac12$ and $X_t(k) \ge \tfrac12$,
		and also $X_t(\ell) < \tfrac12$ by the `else',
		then card $\ell$ was in $\coi{0, \tfrac12}$ initially and card $k$ was not; the swap moves $\ell$ out and $k$ in.
	\end{itemize}
	
	\item 
	Else,
	$k < \tfrac12 \le \ell$ and $X_t(\ell) < \tfrac12 \le X_t(k)$, in which case the count decreases by $1$.
\end{itemize}

We now determine the rate at which the count $M_t$ decreases by $1$.
Suppose that $M_t = m$:
\begin{itemize}
	\item 
	there are $m$ cards with label in $\coi{0, \tfrac12}$ that currently lie outside $\coi{0, \tfrac12}$;
	
	\item 
	there are $m$ cards with label not in $\coi{0, \tfrac12}$ that currently lie in $\coi{0, \tfrac12}$.
\end{itemize}
Comparing a pair of cards one of which is in $\coi{0, \tfrac12}$ and the other not is precisely the case in which $M$ increases.
There are $m^2$ such pairs.
Hence, $M \to M - 1$ at rate $m^2$.
The claimed stochastic domination follows easily from this and the non-increasing property of $M$.
\end{Proof}

This dominates the sorting time of a \emph{given} string by a random variable $T_\star$ that is independent of $n$.
However, in order to sort $n+1$ strings, we need polynomially small failure probability.
An exponential moment on $T_\star$ is easy to establish, providing the following bound.

\begin{lem}
\label{res:unif:mgf-tail}
With the notation from \cref{res:unif:rate-hit},
\[
	\ex{ e^{\theta T_\star} }
\le
	e^{4\theta}
\text{ for all }
	\theta \in [0, \tfrac12]
\Qand
	\pr{ T_\star > 4 \log n }
\le
	e^2 / n^2.
\]
\end{lem}


We now have all the ingredients required to conclude.

\begin{Proof}[Proof of \cref{res:unif:sorting-time}]
A given 0--1 string of length $n$ is sorted in time $4 \log n$ with failure probability $\Oh{1/n^2}$ by \cref{res:unif:rate-hit,res:unif:mgf-tail}.
Hence, any collection of $n+1$ such strings is sorted in time $4 \log n$ with failure probability $\Oh{1/n}$,
as required.
	%
\end{Proof}

%
%

\section{Harmonic Sorter}
\label{sec:harm}

The reader is encouraged to recall the intuition for the harmonic sorter given in \S\ref{sec:outline}.

\begin{defn}
The \textit{harmonic sorter} is the randomised sorter from \cref{def:intro:graph} with
\[
	V
\cq
	[n]
\Qand
	w(\set{i,j})
\cq
	4/\abs{j - i}
\Qfor
	i,j \in [n]
\text{ with }
	i \ne j.
\]
\end{defn}

\begin{rmkt}
We did not include the $4$ in \cref{res:intro:harm}.
This does not affect the result, except scaling time by a factor $4$.
Its inclusion makes the proof slightly cleaner in a few~places.
\end{rmkt}

Let us compare the harmonic sorter with the Gray hypercube, described in \S\ref{sec:outline}.

\begin{itemize}
	\item 
	In the hypercube, a vertex of the form $i \in 00\mstar$ connects to one specific vertex of $11\mstar$, with weight $1$. So, the total comparison rate between $i$ and $11\mstar$ is $w(i, 11\mstar) = 1$.
	
	\item 
	In the harmonic graph, a vertex $i \in 00\mstar$ connects to \emph{all} $n/4$ vertices in $11\mstar$, each with weight in $[1/n, 2/n]$. So, the total weight $w(i, 11\mstar) \in [1/4, 1/2]$, so still $w(i, 11\mstar) \asymp 1$.
\end{itemize}

The analogous unweighted, sparse graph would have $w(i, i + 2^d) = 1$ for all $i \in [n]$ and $d \in [\lg n]$ such that $i + 2^d < n$.
\textcite{I:random-sorting-network-se} suggested a very similar construction, with a slightly different weighting, inspired by the distribution of comparators in the bitonic sort.

\begin{thm}[\cref{res:intro:harm}]
\label{res:harm:sorting-time}
The harmonic sorter
requires order $n (\lg n)^2$ comparisons to sort any length-$n$ list,
both in expectation and with polynomially high probability.
\end{thm}

We split the proof of the upper bound into two parts, for pedagogical reasons.
\begin{enumerate}
	\item 
	Sorting time order $N^2 = (\lg n)^2$ is proved.
	
	\item 
	The method is refined further, achieving the true $N = \lg n$ order.
\end{enumerate}
Prior to that,
we prove a preliminary result on the graph and the lower bound.
Finally, we explain how to adjust the proofs to handle the parallel frameworks of \cref{res:intro:par:n,res:intro:par:p}.

\subsection{Graph Preliminaries and Lower Bound}
\label{sec:harm:prelim}

Before starting the proof, we note that $w(E) \asymp n \lg n$; see \cref{res:harm:prelim:total-weight}.
The proof will be conducted in continuous time, with pair $(i,j)$ compared at rate $w(i,j) = 4/\abs{j-i}$, with $i \ne j$.
Thus, we need to show the continuous-time version sorts in time order~$N = \lg n$.


\begin{lem}
\label{res:harm:prelim:total-weight}
The total weight $w(E)$ satisfies
\[
	n \log n
-	n
\le
	\tfrac14 w(E)
\le
	n \log n
+	n.
\]
\end{lem}

\begin{Proof}
This follows from some basic algebraic manipulations and the inequalities
\[
	\log(m + 1)
\le
	\sumt{d=1}[m]
	1/d
\le
	\log m + 1
\Quad{valid for all}
	m \ge 1.
\qedhere
\]
\end{Proof}

%

To justify the claim that order $N$ is optimal, we give the simple lower bound immediately.

\begin{Proof}[Proof of \cref{res:harm:sorting-time}: Lower Bound]
Consider sorting the `alternating' sequence
\[
	X_0
=
	(2, 1, 4, 3, 6, 5, ..., n, n-1).
\]
The only comparison that can ever do anything is of the form $(2k, 2k+1)$, with $k \in [n/2]$.
Moreover, the sequence is not sorted until \emph{all} $n/2$ such comparisons have been made.
Each happens at rate $4$, in continuous time.
Hence, time $\tfrac18 \log n$ is required both in expectation and with super-polynomially high probability, via a standard coupon-collector argument.
\end{Proof}

Recall that we only consider 0--1 sequences whose length is a power of $2$ and are \textit{balanced}---ie, have the same number of 0-s as 1-s---by \cref{res:prelim:reduction}.

\subsection{Approach 1: Time Order $(\lg n)^2$}
\label{sec:harm:1}

We first establish a bound on the sorting time of order $N^2 = (\lg n)^2$.

\begin{thm}
\label{res:harm:1:main}
The continuous-time harmonic sorter
requires time order $N^2$ to sort any balanced, 0--1-valued list of length $2^N$,
both in expectation and with high probability.
\end{thm}

\begin{Proof}[Strategy]
\qedtriangle
The strategy is not to move a misplaced 0 \emph{directly} into $\coi{0, \tfrac12}$.
Indeed, the long-range edges have low weight, so if there are only, say, 3 misplaced 0-s, then such a move would take \emph{time} order $n$ to happen, as the edge can have weight at most $8/n$.

Instead, the misplaced 0 should be moved closer and closer to $\coi{0, \tfrac12}$, \emph{roughly} halving its distance in every step.
As it gets closer to $\coi{0, \tfrac12}$, the number of 1-s with which it can swap gets smaller, but this is balanced (perfectly) by the increased weight of the short-range~edges.

More concretely, we first `clean' the entirety of $\coi{0, \tfrac14}$/$\coi{\tfrac34, 1}$ of 1-s/0-s, respectively.
Once this is the case, no comparisons with one endpoint in $\coi{0, \tfrac14} \cup \coi{\tfrac34, 1}$ ever leads to a swap, so it is just like sorting a half-length sequence.
We iterate, `cleaning' $\coi{\tfrac14, \tfrac38}$ and $\coi{\tfrac58, \tfrac34}$.
Once we have iterated $N$ times, the entirety of $\coi{0, \tfrac12}/\coi{\tfrac12, 1}$ is clean of 1-s/0-s: the sequence is sorted.

The repeated halving (on both sides) allows each iterate to take time order $\log n \asymp N$, with a uniform implicit constant.
Hence, the $N$ iterations take time order $N^2$ in total.
\end{Proof}

%
%
%

We now proceed formally, with precise definitions and statements.

\begin{defn}
\label{def:harm:1:cuts}
For $k \ge 1$, define
\begin{alignat*}{2}
	I_k^0
&\cq
	\coi{ \tfrac12 - 1/2^k, \tfrac12 - 1/2^{k+1} },
&\quad
	I_k^1
&\cq
	\coi{ \tfrac12 + 1/2^{k+1}, \tfrac12 + 1/2^k };
\\
	I_{< k}^0
&\cq
	\coi{0, \tfrac12 - 1/2^k},
\quad
	I_{\ge k}^0
\cq
	\coi{ \tfrac12 - 1/2^k, \tfrac12 },
&\quad
	I_{< k}^1
&\cq
	\coi{ \tfrac12 + 1/2^k, 1 },
\quad
	I_{\ge k}^1
\cq
	\coi{ \tfrac12, \tfrac12 + 1/2^k };
\\
	I_{< k}
&\cq
	I_{< k}^0 \cup I_{< k}^1,
&\quad
	I_{\ge k}
&\cq
	I_{\ge k}^0 \cup I_{\ge k}^1
=
	\coi{ \tfrac12 - 1/2^k, \tfrac12 + 1/2^k }.
\end{alignat*}
Write $\perm n$ for the set of permutations on $\coi{0,1}_n$.
For $r \ge 0$,
define $\Omega_r \cq \Omega_r^0 \cap \Omega_r^1$
where
\[
\begin{aligned}
	\Omega_r^0
&\cq
	\set*{ x \in \perm n \given x^0 \subseteq \coi{0, \tfrac12} \cup I_{\ge r}^1 }
\\
	\Omega_r^1
&\cq
	\set*{ x \in \perm n \given x^0 \subseteq \coi{\tfrac12, 1} \cup I_{\ge r}^0 }
\end{aligned}
\Qwhere
\begin{aligned}
	x^0
&\cq
	\set*{ i \in \coi{0,1}_n \given x(i) = 0 },
\\
	x^1
&\cq
	\set*{ j \in \coi{0,1}_n \given x(j) = 0 }.
\end{aligned}
\]
	%
\end{defn}

\begin{lem}
\label{res:harm:1:absorbing}
For all $r \ge 0$,
both $\Omega_r^0$ and $\Omega_r^1$ are absorbing sets for the dynamics.
\end{lem}

\begin{Proof}
We show that $\Omega_r^0$ is absorbing; the proof for $\Omega_r^1$ is analogous.

The only way for $X$ to exit $\Omega_r^0$ is if a card with label $\ell < \tfrac12$, which has position less than $\tfrac12 + 1/2^{r+1}$, by assumption, is compared with a card with label $\ell' < \ell$ and position at least $\tfrac12 + 1/2^{r+1}$.
But, all cards with such a position have label $k \ge \tfrac12 > \ell$, by assumption.
\end{Proof}

The time to increase the `cut level' from $r-1$ to $r$ is controlled by the \textit{coupon collector}:
the time to move $n$ cards (`collect $n$ coupons'), where each card is moved at rate $1$ independently.

\begin{defn}
\label{def:harm:cc}
Let $E_m \sim \Exp(m)$ independently for $m \ge 1$.
Then, the \textit{coupon-collector time} for $n \ge 1$ coupons
\(
	\sum_{m=1}^n
	E_m;
\)
denote its law $\mfC(n)$.
\end{defn}


\begin{prop}
\label{res:harm:1:hit}
Let $r \ge 1$ and $X_0 \in \Omega_{r-1}$.
Then,
the hitting time $\tau(\Omega_r)$ of $\Omega_r$ satisfies
\[
	\tau(\Omega_r)
\lesssim
	\mfC_0 + \mfC_1
\Qwhere
	\mfC_0, \mfC_1 \sim^\iid \mfC(n).
\]
\end{prop}

\begin{Proof}
Observe that $\tau(\Omega_r) = \tau(\Omega_r^0) \vee \tau(\Omega_r^1)$.
We analyse the hitting time $\tau(\Omega_r^0)$ of $\Omega_r^0$ first.

Since $\Omega_{r-1}$ is absorbing, by \cref{res:harm:1:absorbing}, the outer intervals $I_{< r}^0 = \coi{0, \tfrac12 - 1/2^r}$ and $I_{< r}^1 = \coi{\tfrac12 + 1/2^r, 1}$ consist only of 0-s and 1-s, respectively.
Hence, the central interval $I_{\ge r} = \coi{ \tfrac12 - 1/2^r, \tfrac12 + 1/2^r }$ contains an equal number of 0-s and 1-s:
	$n/2^r$ of each.
Decomposing
\[
	I_{\ge r}
=
	I_r^0 \cup I_{\ge r+1}^1
\cup
	I_{\ge r+1}^1 \cup I_r^1,
\]
we deduce that there are at least $n/2^{r+1}$ 1-s in
$I_{\ge r} \setminus I_r^1$,
and similarly for 0-s:
\[
	\min\set*{ 
		\abs{ X_t^1 \cap I_{\ge r} \setminus I_r^1 },
	\:	\abs{ X_t^0 \cap I_{\ge r} \setminus I_r^0 }
	}
\ge
	n/2^{r+1},
\]
where $X_t^{0/1}$ is the set of locations of the $0$-s/$1$-s at time $t$, respectively.

Any 1 in $I_r$ can be swapped with any 0 in $I_{\ge r} \setminus I_r$, and such a swap is across distance at most $n/2^{r-1}$.
A given comparison at distance $d$ is performed at rate $4/d$, so such a 1 makes \emph{some} such swap at rate at least
\(
	(n / 2^{r+1}) \cdot 4/(n / 2^{r-1})
=
	1.
\)
Hence, $\tau(\Omega_r^0)$ is upper bounded by the coupon-collector time for $n/2^{r+1} \le n$ coupons, each collected independently at rate~$1$.

Finally, $\tau(\Omega_r^0)$ and $\tau(\Omega_r^1)$ are not independent.
However, each is dominated by $\mfC(n)$, regardless of the initial condition---provided it is in $\Omega_{r-1}$.
Hence, we can upper bound $\tau(\Omega_r) = \tau(\Omega_r^0) \vee \tau(\Omega_r^1)$ by the sum of two independent $\mfC(n)$-s:
	first hit $\Omega_r^0$, then $\Omega_r^1$.
\end{Proof}

We now have all the ingredients required to conclude.

\begin{Proof}[Proof of \cref{res:harm:1:main}]
Our objective is to have $X_t^0 = \coi{0, \tfrac12}$ and $X_t^1 = \coi{\tfrac12, 1}$: ie, hit $\Omega_\infty$.
But, interval $I_k^{0/1}$ contains $\floor{n/2^k}$ positions, so $I_k^{0/1} = \emptyset$ whenever $k > N = \lg n$.
So, it is enough to iterate $N$ times.
Hence,
\(
	\tau_\star
\)
is stochastically dominated by a length-$2N$ sum of iid $\mfC(n)$-s,
by \cref{res:harm:1:hit}.
Each individually has mean order $N = \lg n$ and an exponential tail, thus the sum is order $N^2$ with probability $1 - \Oh{2^{-2N}} = 1 - \Oh{1/n^2}$.
\end{Proof}

\subsection{Approach 2: Time Order $\lg n$}
\label{sec:harm:2}

We improve the bound from $(\lg n)^2$ to $\lg n$ in this subsection.

\begin{thm}
\label{res:harm:2:main}
The continuous-time harmonic sorter
requires time order $N$ to sort any balanced, 0--1-valued list of length $2^N$,
both in expectation and with high probability.
\end{thm}

We reflect on the previous proof, and where it can be improved.
In short, there is no need to wait until \emph{all} of $\coi{0, \tfrac14}$/$\coi{\tfrac34, 1}$ is cleaned before starting cleaning $\coi{\tfrac14, \tfrac38}$/$\coi{\tfrac58, \tfrac34}$.

\begin{Proof}[Reflection]
\qedtriangle
The previous approach was an iterative halver.
\begin{itemize}
	\item 
	Initially, all of $\coi{0, \tfrac12}$ and $\coi{\tfrac12, 1}$ was `dirty' with misplaced 1-s and 0-s, respectively.
	
	\item 
	We cleaned the entirety of $I_1^0 = \coi{0, \tfrac14}$ and $I_1^1 = \coi{\tfrac34, 1}$.
	
	\item 
	Once this was complete,
	we cleaned the entirety of $I_2^0 = \coi{\tfrac14, \tfrac58}$ and $I_2^1 = \coi{\tfrac58, \tfrac34}$.
	
	\item 
	We continued like this until we reached the final level $N$.
\end{itemize}
Now, we ask ourselves:
	``Do we really need \emph{all} the current interval $I_{r-1}$ to be clean before we start cleaning the next interval $I_r$?''
One log factor came from the coupon-collector bound of moving \emph{all} misplaced cards in $I_{r-1}$.
If we allowed ourselves a 1\% tolerance, then the expected time would be upper bounded by $5$, saving a log factor:
\[
	\sumt{m=n/100}[n]
	1/m
\le
	5
\Quad{but}
	\sumt{m=1}[n]
	1/m
\approx
	\log n.
\]

The difficulty now is that the number of 0-s in a certain $I_k^1$ may \emph{increase}: one can come from further to the right.
The idea is that if the density of 0-s in $\coi{\tfrac12, 1}$ decays sufficiently quickly in the distance to $\tfrac12$, then this addition is still outweighed by the 0-s moving out to the left, in the style of \cref{res:harm:1:hit}.
Naturally, this needs careful analysis.
\end{Proof}

Recall the intervals defined in \cref{def:harm:1:cuts}.

\begin{defn}
\label{def:harm:2:cuts}
Let $x \in \perm n$.
We consider the number of `misplaced' 0-s and 1-s:
\begin{alignat*}{2}
	M_k^0(x)
&\cq
	\abs{ x^0 \cap I_k^1 },
&\qquad
	M_k^1(x)
&\cq
	\abs{ x^1 \cap I_k^0 };
\\
	M_{< k}^0(x)
&\mathbin{\eqmakebox[cq][r]{\ensuremath\cq}}
	\abs{ x^0 \cap I_{< k}^1 }
&\qquad
	M_{< k}^1(x)
&\mathbin{\eqmakebox[cq][r]{\ensuremath\cq}}
	\abs{ x^1 \cap I_{< k}^0 }
\\
&\mathbin{\eqmakebox[cq][r]{\ensuremath=}}
	\sumt{\ell=1}[k-1]
	M_k^0(x),
&\qquad
&\mathbin{\eqmakebox[cq][r]{\ensuremath=}}
	\sumt{\ell=1}[k-1]
	M_k^1(x);
\\
	M_k(x)
&\cq
	M_k^0(x) + M_k^1(x),
&\qquad
	M_{< k}(x)
&\cq
	M_{< k}^0(x) + M_{< k}^1(x).
\end{alignat*}
Eg, in words, $M_{< k}^0(x)$ is the number of misplaced 0-s in the interval $I_{< k}^1$.
\end{defn}

In an ordered sequence, $M_k^0(x) = M_k^1(x) = 0$ for all $k$.
Moreover, the dynamics cannot increase the number of 0-s/1-s to the right/left of a given threshold, respectively.
Thus,
\[
	t
\mapsto
	M_{< k}^0(X_t)
\Qand
	t
\mapsto
	M_{< k}^1(X_t)
\quad
	\text{are (deterministically) non-increasing}.
\]

As in the outline, we want to move `most' of the `misplaced' 0-s/1-s in $I_k^1$/$I_k^0$ out in a single round.
We need to take into account the fact that 0-s/1-s from the right/left of $I_k^1$/$I_k^0$ may enter the interval.
Our next result controls the expected number in each interval.

\begin{prop}
\label{res:harm:2:rr:derivation}
For $r \in \mbn$ and $\mft \ge 0$, let $M_k^r \cq M_k(X_{r\mft})$ and $m_k^r \cq \ex{M_k^r}$.
Then,
\[
	m_k^r
\le
	e^{-\mft}
	m_k^{r-1}
+	2
	\sumt{\ell=1}[k-1]
	m_\ell^{r-1}
\Qforall
	r,k \ge 1.
\]
\end{prop}

\begin{Proof}
Suppose that the current state $X_t = x$, and drop $x$ from the notation; eg, write $M_k^0$ for $M_k^0(x)$.
We want to lower bound the rate at which 0-s/1-s in $I_k^1$/$I_k^0$ are moved out.

We start with the 0-s in $I_k^1$; the 1-s in $I_k^0$ are handle analogously.
There are two key aspects in bounding the rate, for a given 0:
\begin{itemize}[noitemsep]
	\item 
	the \emph{number} of 1-s in the `target' set $\coi{0, 1} \setminus I_{\le k}^1$;
	
	\item 
	the \emph{distance} at which these 1-s lie.
\end{itemize}
In order that the distance is not too large, we consider comparisons inside $I_{\ge k} = \coi{ \tfrac12 - 2^{-k}, \tfrac12 + 2^{-k} }$; these always have distance at most $2 \cdot 2^{-k} n = 2^{-k+1} n$.

If the permutation $x$ were sorted, then there would be precisely half the positions in $I_{\ge k}$ would hold a 1.
However, for every 1 in $I_{< k}^0$ there may be an extra 0 in $I_{\ge k}$.
Hence,
\[
	\abs{ x(.0\star) \cap I_{\ge 3} }
\ge
	\tfrac12
	\abs{ I_{\ge k} }
	n
-	\abs{ x(.1\star) \cap I_{< k}^0 }
=
	(2^{-k} - M_{< k}^1/n) n.
\]
Thus, the number of 1-s inside $I_{\ge k} \setminus I_k^1$ is at least
\[
	(2^{-k} - 2^{-(k+1)} - M_{< k}^1/n) n
=
	(2^{-k-1} - M_{< k}^1/n) n,
\]
since there only $2^{-(k+1)} n$ positions in $I_k^1$, and each is a distance at most $2^{-k+1} n$ away.
Thus, \emph{each} 0 in $I_k^1$ moves out independently at rate at least
\[
	(2^{-k-1} - M_{< k}^1/n) \cdot 4 / 2^{-k+1}
=
	1 - 2 M_{< k}^1.
\]
We emphasise that the rate at which the \emph{0}-s decrease depends on the positions of the \emph{1}-s.

An analogous argument applies to moving the 1-s out of $I_k^0$.
Hence,
\[
	M_k \to M_k - 1
\Quad{at rate at least}
	2 (1 - M_{< k}/n) M_k
\ge
	M_k
\Qif
	M_{< k}/n
\le
	\tfrac12.
\]

If we know that $M_k \to M_k - 1$ at rate at least $M_k$, we want to know how many 0-s/1-s remain misplaced after time $\mft$.
This (lower bound) is equivalent to the standard coupon-collector problem, in which coupons are collected independently at rate 1.
The probability a given coupon \emph{is not} collected by time $\mft$ is simply $\pr{ \Exp(1) > \mft } = e^{-\mft}$, independently.
Hence, the number of coupons which are collected in time $\mft$ is lower bounded by a length-$M_k^{r-1}$ sum $B$ of $\Bern(e^{-\mft})$-s; in fact, $B \sim \Bin(M_k^{r-1}, e^{-\mft})$, by the independence.

On the complementary event that $M_{< k}/n > \tfrac12$, we assume that no 0-s/1-s are moved out of $I_k^1$/$I_k^0$ in the round; ie, we lower bound the number by $0$.

Whilst $t \mapsto M_{< k}(X_t)$ is non-increasing, it is not the case for $t \mapsto M_k(X_t)$. Indeed, a misplaced 0 may move out of $I_{k-1}^1$ into $I_k^1$, decreasing $M_{k-1}^0$ but increasing $M_k^1$.
We upper bound the number of such 0-s/1-s which enter $I_k^1$/$I_k^0$ during round $r$---ie, time $\coi{ (r-1) \mft, r \mft }$---by the total number $M_{< k}^0$/$M_{< k}^1$ above/below at the start of the round.

Putting all this together, we obtain the following stochastic upper bound:
\[
	M_k^r
&
\lesssim
	\one{ M_{< k}^{r-1}/n \le \tfrac12 }
	B
+	\one{ M_{< k}^{r-1}/n > \tfrac12 }
	M_k^{r-1}
+	M_{< k}^{r-1}
\\&
\lesssim
	B
+	\one{ M_{< k}^{r-1}/n > \tfrac12 } n
+	M_{< k}^{r-1},
\]
since $M_k^{r-1} \le n$ deterministically.
Taking expectation, and applying Markov's inequality,
\[
	\ex{ M_k^r }
&
\le
	e^{-\mft}
	\ex{ M_k^{r-1} }
+	\pr{ M_{< k}^{r-1} > \tfrac12 n } n
+	\ex{ M_{< k}^{r-1} }
\\&
\le
	e^{-\mft}
	\ex{ M_k^{r-1} }
+	3
	\ex{ M_{< k}^{r-1} }
=
	e^{-\mft}
	\ex{ M_k^{r-1} }
+	\sumt{\ell=1}[k-1]
	M_\ell^{r-1}.
\qedhere
\]
\end{Proof}

Now that we have this relation on the expectations, we need to bound the solution.

\begin{lem}
\label{res:harm:2:rr:bound}
Suppose that $((m_k^r)_{k=1}^N)_{r\ge0}$ satisfies $m_k^r \le n$ for all $k,r \ge 0$ and
\[
	m_k^r
\le
	\tfrac13
	m_k^{r-1}
+	2
	\sumt{\ell=1}[k-1]
	m_\ell^{r-1}
\Qforall
	r,k \ge 1.
\]
Then,
\[
	m_k^r
\le
	2^{-(r - 3k)} n
\Qforall
	r, k \ge 0.
\]
\end{lem}

\begin{Proof}
We prove this by induction on $r$.
The base case $r = 0$ is immediate, as $m_k^r \le n$.

Now, assume that the inequality holds for $r-1$ (and all $k$).
We prove it for $r$ (and all $k$):
\[
	m_k^r
\le
	\tfrac13
	m_k^{r-1}
+	2
	\sumt{\ell=1}[k-1]
	m_\ell^{r-1}
&
\le
	2^{-(r-1 - 3k)}
	\rbb{ 
		\tfrac13
	+	\sumt{\ell=1}[k-1]
		2^{3(k-\ell)}
	}
	n
\\&
\le
	2^{-(r - 3k)}
\cdot
	2
	\rbb{ 
		\tfrac13
	+	\sumt{\ell=1}[\infty]
		8^{-\ell}
	}
	n
\le
	2^{-(r - 3k)}
	n.
\qedhere
\]
\end{Proof}

%

It is easy to combine the previous two results to find the halving time.


\begin{Proof}[Proof of \cref{res:harm:2:main}]
Take $\mft \cq 2 \ge \log 3$, so that $e^{-\mft} \le \tfrac13$.
Then,
\[
	\ex{ M_k(X_{2r}) }
\le
	2^{-(r - 3k)} n
\Qforall
	r,k \ge 1,
\]
by \cref{res:harm:2:rr:derivation,res:harm:2:rr:bound} together.
Now, $k \le N$, else $I_k = \emptyset$, to take $r \cq 7 N \ge 6 N + \lg N$.
Then, applying Markov's inequality,
noting that $M_k(X_{2r}) \in \mbn$,
\[
	\pr{ M_k(X_{2r}) \ne 0 }
\le
	\ex{ M_k(X_{2r}) }
\le
	2^{-(r-3k)} n
\le
	2^{-2N} / N
\Qforall
	k \ge 1,
\]
as $n = 2^N$.
A union bound over the $N$ values of $k$ completes the proof:
\[
	\pr{ \tau_\star > 14 N }
\le
	2^{-2N}
=
	1/n^2.
\qedhere
\]
\end{Proof}

\subsection{Parallel Comparisons}
\label{sec:harm:par}

We now consider a discrete-time, synchronised, parallel approach:
	in each step, a disjoint collection $\set{(i_\ell, j_\ell)}_{\ell=1}^L$ of pairs is chosen;
	the labels at positions $i_\ell$ and $j_\ell$ are sorted, for each $k \in [L]$.
The number $L$ of pairs in a collection may be random.

Sorting a collection of pairs simultaneously breaks the independence between different pairs provided by the continuous-time nature of the previous proof.
We show below that the proof is actually very robust to this, and only minor adjustments are required to sort multiple pairs simultaneously, in discrete time.
All that really matters is that the probability that a particular pair $\set{i,j}$ is chosen is lower bounded by a constant times $1/\abs{j - i}$.

\begin{defn}
Let $\mcm$ denote the set of \textit{matchings} in $[n]$:
	the set of all collections of disjoint pairs.
For $\alpha > 0$, let $\mcq_\alpha$ denote the set of probability measures $\tilde q$ on $\mcm$ such that
\[
	q_{i,j}
\cq
	\pr[M\sim \tilde q]{ \set{i,j} \in M }
=
	\sumt{M \in \mcm : \set{i,j} \in M}
	\tilde q(M)
\ge
	\alpha / \abs{j - i}
\Qforall
	i,j \in [n]
\text{ with }
	i \ne j.
\]
\end{defn}

The harmonic sorter is an example of such a distribution---a single edge is a matching, trivially---with $\alpha \asymp 1/(n \log n)$, since $w(E) \asymp n \log n$ by \cref{res:harm:prelim:total-weight}.
The idea is that increasing $\alpha$ speeds up the selection process.
So, the final sorting time is order $(\log n) / \alpha$.

\begin{prop}
\label{res:harm:par:rr:derivation}
Let $n \in \mbn$ and let $\alpha \in (0,1)$, allowed to depend on $n$.
Suppose that $\tilde q \in \mcq_\alpha$, and let $X_t$ denote the state of the list after $t$ matchings drawn from $\tilde q$ have been processed ($t \ge 0$).
For $r \in \mbn$, let $M_k^r \cq M_k(X_{5 r / \alpha})$ and $m_k^r \cq \ex{M_k^r}$.
Then,
\[
	m_k^r
\le
	\tfrac13
	m_k^{r-1}
+	2
	\sumt{\ell=1}[k-1]
	m_\ell^{r-1}
\Qforall
	r,k \ge 1.
\]
	%
\end{prop}

\begin{Proof}
The events that the pair $\set{i,j}$ appears in a (sequence of) matchings is not independent across different pairs $\set{i,j}$, unlike in the continuous-time case.
Recalling the notation from \cref{def:harm:2:cuts},
in the proof of \cref{res:harm:2:rr:derivation}, we showed that
\[
	M_k^r
\lesssim
	B
+	\one{ M_{< k}^{r-1}/n > \tfrac12 } n
+	M_{< k}^{r-1},
\]
where $B$ was a length-$M_k^{r-1}$ sum of $\Bern(e^{-\mft})$-s.
The only important difference now is that the Bernoullis are not independent. But, we only used the expected value of $B$, which does not require independence of the summands.
We also need to check the time scaling.

To explain this further, if we ran for a very short time $\delta$, then the probability that some pair is picked causing $M_k \to M_k - 1$ \emph{was} (lower bounded by) approximately $\delta$, where pair $\set{i,j}$ was picked with probability approximately $4 \delta / \abs{j - i}$.
If we consider $5 \delta / \alpha$ iid matchings, then the probability that $\set{i,j}$ is chosen in (at least) one of these is at least
\[
	1 - \rbb{ 1 - \alpha / \abs{j - i} }{}^{5 \delta / \alpha}
\ge
	4 \delta / \abs{j - i}
=
	\delta w(i,j).
\]
So, $5 \delta / \alpha$ iid matchings are like $\delta$ units of continuous time.

We only need one pair to be picked to cause $M_k \to M_k - 1$.
So, by the same arguments as in \cref{res:harm:2:rr:derivation} and the union bound, there is probability $\delta$ of this happening in $5 \delta / \alpha$ iid matchings.
Repeating this $1/\delta$ times, the probability in the Bernoullis is at most $\tfrac13$.
\end{Proof}

We can now deduce the sorting time when $\tilde q \in \mcq_\alpha$.

\begin{prop}
\label{res:harm:par:sorting-gen}
Let $n \in \mbn$ and let $\alpha \in (0,1)$, allowed to depend on $n$.
Let $\tau_\star$ denote the sorting time of the associated parallel sorter, started from an arbitrary $0$--$1$ sequence.
Then,
\[
	\pr{ \tau_\star > 50 (\log n) / \alpha }
\le
	1/n^2.
\]
\end{prop}

\begin{Proof}
Applying \cref{res:harm:2:rr:bound} to \cref{res:harm:par:rr:derivation},
\[
	\ex{ M_k(X_{5 r / \alpha}) }
\le
	2^{-(r - 3k)} n.
\]
Recall that $N = \lg n$.
Taking with $r \ge \tfrac{34}5 N = (6 + \tfrac45) N \ge 6 N + \lg N$, we deduce that
\[
	\pr{ M_k(X_{34 (\lg n) / \alpha}) \ne 0 }
\le
	2^{-2N} / N.
\]
The union bound over $k$, along with the fact that $34 \lg n \le 50 \log n$, then gives
\[
	\pr{ \tau_\star > 50 (\log n) / \alpha }
\le
	2^{-2N}
=
	1/n^2.
\qedhere
\]
\end{Proof}

Finally, we must show that the constructions from \cref{res:intro:par:n,res:intro:par:p} lie in some~$\mcq_\alpha$.


\begin{lem}
\label{res:harm:par:tilde-q:n}
The distribution $\tilde q$ in \cref{res:intro:par:n} satisfies $\tilde q \in \mcq_\alpha$ for $\alpha \cq 1/(4 \lg n)$.
\end{lem}

\begin{Proof}
Fix a pair $\set{i,j}$ with $\abs{i - j} = d$.
Given $D = d$, all edges of length $d$ lie in $M_0 \cup M_1 \cup M_2 \cup M_3$;
so, there is a probability $1/4$ of choosing $\set{i,j}$.
So, $\set{i,j}$ is chosen with probability $\tfrac14 \pr{D = d}$.
Let $k = \floor{\lg(n/d)}$, so that $2^{k+1}/n < d \le 2^k/n$.
Direct calculation gives
\[
	\pr{ D = d }
&
=
	\pr{ D = d \given K = k } \pr{ K = k }
\\&
=
	2^{k+1}/n \cdot 1/\lg n
\ge
	2^{\lg(n/d)} / (n \lg n)
=
	1 / (d \lg n).
\qedhere
\]
\end{Proof}

\begin{lem}
\label{res:harm:par:tilde-q:p}
The distribution $\tilde q$ in \cref{res:intro:par:p} satisfies $\tilde q \in \mcq_\alpha$ for $\alpha \cq 1/\rbb{ 4(n/p) \log n }$.
\end{lem}

\begin{Proof}
In order to sort $\set{i,j}$, the pair must be chosen and both $i$ and $j$ never chosen again.

\begin{itemize}
\item \textsl{At least once.}
If $\set{I,J} \sim w$, then, by \cref{res:harm:prelim:total-weight},
\[
	\pr{ \set{I,J} = \set{i,j} }
=
	w(\set{i,j}) / w(E)
\ge
	1 / \rbb{ \abs{j - i} \cdot n \log(ne) }.
\]
So, the probability that $\set{i,j}$ is amongst $p$ iid samples is at least
\[
	1 - \rbb{ 1 - 1 / \rbb{ \abs{j - i} \cdot n \log(ne) } }{}^p
&
\ge
	1 - \exp*{ p / \rbb{ \abs{j - i} n \log(ne) } }
\\&
\ge
	\rbr{ 2 (n/p) \log n }^{-1} / \abs{j - i}.
\]

\item \textsl{At most once.}
By symmetry, the endpoints of the edge are both \emph{marginally} uniform.
So, by the union bound, the probability that a single sample contains any given vertex is at most $2/n$, and the probability that any amongst $p$ samples is at most $2p/n \le \tfrac12$.
\end{itemize}
Combining these considerations, we may take $\alpha \cq 1 / \rbb{ 4 (n/p) \log n }$.
\end{Proof}


\cref{res:intro:par:n,res:intro:par:p} follow easily from \cref{res:harm:par:sorting-gen,res:harm:par:tilde-q:n,res:harm:par:tilde-q:p}.

\renewcommand{\bibfont}{\sffamily\small}
\renewcommand{\bibfont}{\sffamily}
\printbibliography[title = {Bibliography}]

\appendix
\section{Appendix}

In this appendix, we give two alternative proofs of the order $n^2 \log n$ sorting time for the uniform sorter.
To recall, this sorts the elements at a uniformly chosen pair of locations.

We use continuous time here, selecting each pair at rate $1$. So, the $n^2 \log n$ discrete-time bound becomes an order $\log n$ bound in continuous time, since there are $\binom n2 \asymp n^2$ pairs.

\subsection{Recursive Argument}

Applying the 0--1 principle makes the intuition a little fuzzy.
We also give a recursive argument now, which makes it clearer how the algorithm refines the order.

\begin{nota}
Represent cards and positions by their fractional length-$N$ binary representation.
Eg, if $n = 2^3$, then $n/2 \rightsquigarrow .1$, $n/4 \rightsquigarrow .01$, $3n/4 \rightsquigarrow .11$, etc.
For $b \in \set{0,1}^\mbn$,~let
\[
	.b\star
\cq
	\set{ .b' \in \set{0,1}^N \given b'_r = b_r \: \forall \, r \le \abs b },
\Qwhere
	\text{$\abs b$ is the length of $b$}.
\]
\end{nota}

\begin{figure}
\centering
\includegraphics[width = 0.9\linewidth]{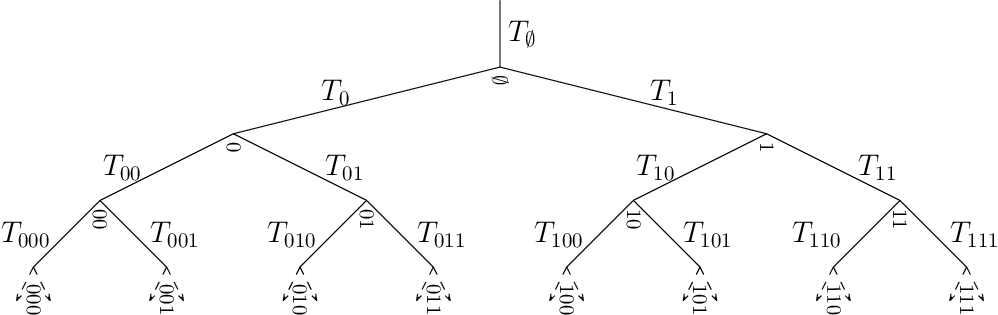}
\caption{Branching process from recursion.
	Vertex labels indicate which binary interval has just been split;
	eg, label $01$ indicates that $X_t(.010\mstar) = .010\mstar$ and $X_t(.011\mstar) = .011\mstar$ at this point.
	Edge labels indicate the passage time;
	eg, label $T_{01}$ indicates that it took time $T_{01}$ between achieving $\set{ X_t(.01\mstar) = .01\mstar }$ and $\set{ X_t(.010\mstar) = .010\mstar } \cap \set{ X_t(.011\mstar) = .011\mstar }$%
	}
\label{fig:unif:binarytree}
\end{figure}

\begin{Proof}[Outline]
\qedtriangle
The recursive argument splits the deck in half in each round.
\begin{itemize}
\item 
Run until all $2^{N-1}$ cards with labels in $0\star$ have positions in $0\star$:
\[
	S_\emptyset
\cq
	\inf\set{ t \ge 0 \given X_t(.0\star) = .0\star }
\Qand
	T_\emptyset
\cq
	S_\emptyset.
\]
At this time, necessarily $X_t(.1\star) = .\emptyset\mstar \setminus X_t(.0\star) = .\emptyset\mstar \setminus .0\star = .1\star$.

\item 
Now, all cards with position in $.0\star$ have label smaller than all those in $.1\star$.
So, any comparison between cards in positions $(i, j)$ with $i < .1 \le j$ after $S_\emptyset$ has no effect.

\item 
Thus, we can recurse; the recursive branching structure is shown in \cref{fig:unif:binarytree}.

\item 
Once the labels have length $N$, they correspond to only a singleton.
After the \textit{last passage time}---ie, the maximal sum along ancestries---all cards are correctly positioned.

\item 
We can stochastically dominate each $T_b$ as in \cref{res:unif:rate-hit}.
A large-deviation bound shows that the maximum over the $2^N$ ancestries is order $\log n$.
\qedhere
\end{itemize}
\end{Proof}

We now make this formal, but do not redo the proofs, instead referring to earlier ones.

\begin{defn}
For $b \in \set{0,1}^\mbn \setminus \emptyset$, write $b' \in \set{0,1}^\mbn$ for $b$ without its last bit
and
\[
	S_b
\cq
	\inf\set{ t \ge S_{b'} \given X_t(.b0\star) = .b0\star }
\Qand
	T_b
\cq
	S_b - S_{b'},
\]
with the convention $S_{\emptyset'} \cq 0$.
See \cref{fig:unif:binarytree} for a pictorial representation.
\end{defn}

\begin{lem}
\label{res:unif:preservation}
If $t \ge S_b$, then $X_t(.b0\star) = .b0\star$ and $X_t(.b1\star) = .b1\star$.
\end{lem}

\begin{Proof}
At time $S_b$, all cards with label $\ell < .b$ have position $i < .b$ and those with label $k \ge .b$ have position $j \ge .b$.
So, any comparison between cards in positions $(i,j)$ with $i < .b < j$ after time $S_b$ has no effect, preserving this property.
\end{Proof}

\begin{defn}
Let
\(
	M_N
\cq
	\max_{b \in \set{0,1}^N} S_b
\)
denote the \textit{last-passage time} for the binary tree.
\end{defn}

First and foremost, we need to relate the last-passage time to the sorting time.

\begin{lem}
\label{res:unif:lpp:sorting}
The sorting time $\tau_\star$ satisfies $\tau_\star \lesssim M_N$.
\end{lem}

\begin{Proof}
This is a simple consequence of the preservation property of \cref{res:unif:preservation}.
	%
\end{Proof}

It suffices to control the last-passage time.

\begin{lem}
\label{res:unif:lpp:bound}
The last-passage time $M_N$ is order $\log n$ in expectation and whp.
\end{lem}

\begin{Proof}
The preservation from \cref{res:unif:preservation} allows the argument of \cref{res:unif:rate-hit} to be applied to each substring.
Hence, $T_b \lesssim T_\star \cq \sum_{m\ge1} E_m$, where $E_m \sim \Exp(m^2)$ independently for $m \ge 1$, recalling that each pair is compared at rate $1$, so no scaling is needed to account for the fact that there are only $n/2^{\abs b}$ cards in $.b\star$.

The last-passage time $M_N$ is a maximum over $2^N$ random variables $S_b$.
Each of these is a upper bounded by sum of $N$ iid random variables $T_\star$, and $T_\star$ has mean order $1$ and an exponential tail, by \cref{res:unif:mgf-tail}.
A simple Chernoff bound now establishes the claim.
\end{Proof}

\begin{Proof}[Proof of \cref{res:intro:unif}]
This now follows immediately from \cref{res:unif:lpp:sorting,res:unif:lpp:bound}.
	%
\end{Proof}

\subsection{Inversion-Based Proof}

\newcommand{\xx}{x}

A simple argument, based on the number of inversions, was outlined by \textcite{M:uniform-sort}.
Whilst this does not give such intuition for how the sorter operates, the proof is short.

We fill in the details here, for completeness, but claim no originality for the idea.

\begin{defn}
Let $\xx$ be a permutation on $[n]$.
Then, an \textit{inversion} is a pair $(i,j)$ with $i < j$ and $\xx(i) > \xx(j)$.
Write $\inv \xx$ for the number of inversions in $\xx$:
\[
	\inv \xx
\cq
	\abs{ \set{ (i,j) \in [n]^2 \given \xx(i) > \xx(j) } }.
\]
	%
\end{defn}

Trivially, $\inv \xx \le n^2$.
The key is that a sorting step decreases the number of inversions.

\begin{lem}
Let $\xx$ be a permutation and $(i,j)$ be an inversion for $\xx$:
$i,j \in [n]$ with $i < j$ and $\xx(i) > \xx(j)$.
Let $\xx'$ denote the permutation resulting from sorting $i$ and $j$:
\[
	\xx'(i)
\cq
	\xx(i) \wedge \xx(j),
\quad
	\xx'(j)
\cq
	\xx(i) \vee \xx(j)
\Qand
	\xx'(k)
\cq
	\xx(k)
\Qfor
	k \notin \set{i,j}.
\]
Then,
\[
	\inv \xx'
\le
	\inv \xx - 1.
\]
\end{lem}

\begin{Proof}
	%
No new inversions are created, after relabelling,
and $(i,j)$ is no longer one.
\end{Proof}

It is simple to estimate how long it takes to sort, establishing \cref{res:intro:unif}.

\begin{prop}
The process $(\inv X_t)_{t\ge0}$ of inversion counts is stochastically dominated by a process $(I_t)_{t\ge0}$ with $I_0 \cq \inv X_0$ and $I_t \to I_t - 1$ at rate $I_t$.
In particular,
\[
	\pr{ \tau_\star > n^2 \log n + \tfrac12 \alpha n^2 }
\le
	e^{-\alpha}
\Qand
	\ex{ \tau_\star }
=
	n^2 \log n + \Oh{n^2}
\asymp
	n^2 \log n,
\]
where $\tau_\star$ is the sorting time, as before.
\end{prop}

\begin{Proof}
If the pair chosen does not form an inversion, then the comparison has no effect on the process; in particular, the number of inversions does not change.
If the pair does form an inversion, then the number of inversions decreases by at least $1$.
Each pair is chosen at rate $1$, hence the claimed stochastic domination holds.
	
We know by the coupon-collector problem that
\[
	\tau_\star
\lesssim
	\binomt n2
	\max\set{ E_1, ..., E_{n^2} }
\quad
	\text{in distribution}
\Qwhere
	E_1, E_2, ... \sim^\iid \Exp(1).
\]
In particular,
\[
	\pr{ \tau_\star / \binomt n2 > \log(n^2) + \alpha }
\le
	n^2 e^{-\log(n^2) - \alpha}
=
	e^{-\alpha}.
\]
But, observe that $\binom n2 \le \tfrac12 n^2$ and $\log(n^2) = 2 \log n$.
This completes the proof.
\end{Proof}

\end{document}

%% file: sorting_preamble_biblio.tex
\usepackage[doi=false,isbn=false,url=false,
	hyperref=auto,
	sorting=nyt,
	maxnames=10,
	maxcitenames=3,
	backend=biber,
	texencoding=auto,
	giveninits=true,
	block=space,
]{biblatex}

\DefineBibliographyStrings{english}{%
	andothers = {et al}
} 
 
\DeclareFieldFormat{eprint:arxiv}{%
	\em \href{https://arxiv.org/abs/#1}{arXiv:\allowbreak #1}}

\DeclareFieldFormat{eprint:mrnumber}{%
	\href{http://www.ams.org/mathscinet-getitem?mr=MR#1}{MR\allowbreak #1}}

\DeclareFieldFormat{eprint:jstor}{%
	\href{http://www.jstor.org/stable/#1}{JSTOR\allowbreak #1}}

\DeclareFieldFormat{eprint:customeprint}{%
	\upshape \url{#1}}

\DeclareFieldFormat{eprint:inprep}{%
	\em In preparation}

\DeclareFieldFormat{eprint:note}{%
	\em {#1}}

\DeclareFieldFormat{eprint:onarxiv}{%
	\em Available on arXiv}

\DeclareFieldFormat{eprint:toappear}{%
	\mbox{\em To appear}}

\DeclareSourcemap{
	\maps[datatype=bibtex]{
		\map{
			\step[fieldsource=mrnumber,		fieldtarget=eprint, final]
			\step[fieldset=eprinttype,		fieldvalue=mrnumber]
		}
		\map{
			\step[fieldsource=arxiv,		fieldtarget=eprint, final]
			\step[fieldset=eprinttype,		fieldvalue=arxiv]
		}
		\map{
			\step[fieldsource=jstor,		fieldtarget=eprint, final]
			\step[fieldset=eprinttype,		fieldvalue=jstor]
		}
		\map{
			\step[fieldsource=customeprint,	fieldtarget=eprint, final]
			\step[fieldset=eprinttype,		fieldvalue=customeprint]
		}
		\map{
			\step[fieldsource=inprep,		fieldtarget=eprint, final]
			\step[fieldset=eprinttype,		fieldvalue=inprep]
		}
		\map{
			\step[fieldsource=note,			fieldtarget=eprint, final]
			\step[fieldset=eprinttype,		fieldvalue=note]
		}
		\map{
			\step[fieldsource=onarxiv,		fieldtarget=eprint, final]
			\step[fieldset=eprinttype,		fieldvalue=onarxiv]
		}
		\map{
			\step[fieldsource=toappear,		fieldtarget=eprint, final]
			\step[fieldset=eprinttype,		fieldvalue=toappear]
		}
	}
}

\DeclareFieldFormat{doi}{%
	\href{https://doi.org/#1}{DOI}%
}

\DeclareFieldFormat{url}{%
	\href{https://#1}{\texttt{#1}}%
}

\DeclareFieldFormat{comments}{%
	\em #1%
}

\DeclareFieldFormat
	[article,inbook,incollection,inproceedings,patent,thesis,unpublished]
	{title}{{#1\isdot}}

\DeclareFieldFormat
	[article,book,inbook,incollection,inproceedings,patent,thesis,unpublished, online]
	{date}{\mkbibparens{#1}}
\DeclareFieldFormat
	[article,book,inbook,incollection,inproceedings,patent,thesis,unpublished]
	{volume}{\mkbibbold{#1}}
\DeclareFieldFormat
	{pages}{\mkbibparens{#1}}

\DeclareFieldFormat{edition}{%
	\ifinteger{#1}
	{\mkbibordedition{#1}~\bibstring{edition}\addcomma}
	{#1~\bibstring{edition}\addcomma}}

\renewbibmacro*{volume+number+eid}{%
	\printfield{volume}%
\setunit*{\adddot}%
	\printfield{number}%
\setunit{\space}%
	\printfield{eid}%
}

\DeclareBibliographyDriver{article}{%
	\usebibmacro{bibindex}%
	\usebibmacro{begentry}%
	\usebibmacro{author}%
\setunit{\addspace}%
	\usebibmacro{date}%
\newunit\newblock
	\usebibmacro{title}%
\newunit\newblock
	\printfield{journaltitle}\addperiod%
\setunit*{\addspace}%
	\usebibmacro{volume+number+eid}%
\setunit*{\addspace}\newblock
	\printfield{pages}
\setunit*{\addspace}
	\printfield{comments}%
	\usebibmacro{eprint}%
\setunit*{\addnbspace}%
	\printfield{doi}%
	\usebibmacro{finentry}%
}

\DeclareBibliographyDriver{online}{%
	\usebibmacro{bibindex}%
	\usebibmacro{begentry}%
	\usebibmacro{author}%
\setunit{\addspace}%
	\usebibmacro{date}%
\newunit\newblock
	\usebibmacro{title}%
\newunit\newblock
	\usebibmacro{eprint}%
	\usebibmacro{finentry}%
}

\DeclareBibliographyDriver{book}{%
	\usebibmacro{bibindex}%
	\usebibmacro{begentry}%
	\usebibmacro{author}%
\setunit{\addspace}\newblock
	\usebibmacro{date}%
\newunit\newblock
	\usebibmacro{title}%
\setunit*{\addspace}\newblock
	\printfield{edition}%
\newunit\newblock
	\printlist{publisher}
\setunit*{\addspace}%
	\printfield{volume}%
\setunit*{\addspace}\newblock%
	\printfield{comments}%
	\usebibmacro{eprint}%
\setunit*{\addspace}
	\printfield{doi}
	\usebibmacro{finentry}%
}

\DeclareBibliographyDriver{thesis}{%
	\usebibmacro{bibindex}%
	\usebibmacro{begentry}%
	\usebibmacro{author}%
\setunit{\addspace}\newblock
	\usebibmacro{date}%
\newunit\newblock
	\usebibmacro{title}.%
\setunit*{\addspace}\newblock
	\space Thesis, \printlist{institution}
	\usebibmacro{eprint}%
	\printfield{comments}%
\setunit*{\addspace}
	\printfield{doi}%
	\usebibmacro{finentry}%
}

\DeclareBibliographyDriver{inproceedings}{%
	\usebibmacro{bibindex}%
	\usebibmacro{begentry}%
	\usebibmacro{author}%
\setunit{\addspace}%
	\usebibmacro{date}%
\newunit\newblock
	\usebibmacro{title}%
\newunit\newblock
	\printfield{booktitle}\addcomma%
\setunit*{\addspace}%
	\printfield{series}\addcomma%
\setunit*{\addspace}\newblock
	\usebibmacro{editor}\addcomma
\setunit*{\addspace}\newblock
	\printlist{publisher}
\setunit*{\addspace}%
	\printfield{volume}%
\setunit*{\addspace}%
	\printfield{pages}
\setunit*{\addspace}
	\usebibmacro{eprint}%
	\printfield{comments}%
\setunit*{\addnnspace}
	\printfield{doi}
	\usebibmacro{finentry}%
}

\DeclareBibliographyDriver{inbook}{%
	\usebibmacro{bibindex}%
	\usebibmacro{begentry}%
	\usebibmacro{author}%
\setunit{\addspace}%
	\usebibmacro{date}%
\newunit\newblock
	\usebibmacro{title}%
\newunit\newblock
	\printfield{booktitle}\addcomma%
\setunit*{\addspace}%
	\printfield{series}\addcomma%
\setunit*{\addspace}\newblock
	\usebibmacro{editor}\addcomma
\setunit*{\addspace}\newblock
	\printlist{publisher}
\setunit*{\addspace}%
	\printfield{volume}%
\setunit*{\addspace}%
	\printfield{pages}
\setunit*{\addspace}
	\usebibmacro{eprint}%
	\printfield{comments}%
\setunit*{\addspace}
	\printfield{doi}
	\usebibmacro{finentry}%
}

\DeclareBibliographyDriver{incollection}{%
	\usebibmacro{bibindex}%
	\usebibmacro{begentry}%
	\usebibmacro{author}%
\setunit{\addspace}%
	\usebibmacro{date}%
\newunit\newblock
	\usebibmacro{title}%
\newunit\newblock
	\printfield{booktitle}\addcomma%
\setunit*{\addspace}%
	\printfield{series}\addcomma%
\setunit*{\addspace}\newblock
	\printlist{publisher}
\setunit*{\addspace}%
	\printfield{volume}%
\setunit*{\addspace}%
	\printfield{pages}
\setunit*{\addspace}
	\usebibmacro{eprint}%
	\printfield{comments}%
\setunit*{\addspace}
	\printfield{doi}
	\usebibmacro{finentry}%
}

\AtEveryBibitem{%
	\clearfield{day}%
	\clearfield{month}%
} 
 
\DeclareNameAlias{sortname}{given-family}
\DeclareNameAlias{default}{given-family}

%% file: sorting_preamble_main.tex
\usepackage[hidelinks]{hyperref}
\hypersetup{
	colorlinks,
	linkcolor={blue},
	citecolor={ForestGreen},
	urlcolor={blue}
}

\usepackage[UKenglish]{babel}
\usepackage{lmodern}
\usepackage[T1]{fontenc}
\usepackage{amsmath, amsthm, amssymb, mathrsfs, bm, mathtools}
\usepackage{wasysym}
\allowdisplaybreaks

\usepackage{titlesec}
\usepackage{xifthen}
\usepackage{xspace}

\usepackage{xparse}

\usepackage{graphicx}
\usepackage[font = normal, labelfont = {bf, sf}, labelsep = period, singlelinecheck = false]{caption}
\numberwithin{figure}{section}
\numberwithin{table}{section}

\usepackage{titlesec}
\newcommand{\fullstopafter}[1]{#1.}
\titleformat{\section}
	{\sffamily \Large \bfseries \boldmath}{\thesection}{1em}{}
\titleformat{\subsection}
	{\sffamily \large \bfseries \boldmath}{\thesubsection}{1em}{}
\titleformat{\subsubsection}[runin]
	{\sffamily \normalsize \bfseries \boldmath}{\thesubsubsection}{1em}{\fullstopafter}
\titleformat{\paragraph}[runin]
	{\normalsize \itshape}{\theparagraph}{1em}{\fullstopafter}
\titlespacing*{\paragraph}
	{0pt}{\medskipamount}{*2.5}
\titleformat{\subparagraph}[runin]
	{\normalsize \itshape}{\thesubparagraph}{1em}{\fullstopafter}

\usepackage{enumitem}

\setlist[enumerate]{%
	topsep	= \smallskipamount,
	itemsep = 0pt,
}
\setlist[itemize]{%
	topsep	= \smallskipamount,
	itemsep	= 0pt,
	label	= \bcdot
}
\setlist[description]{%
	topsep	= \medskipamount,		
	itemsep	= \smallskipamount,		
	font	= {\mdseries\slshape},	
}

\newcommand{\negphantom}[1]{\ifmmode\settowidth{\dimen0}{$#1$}\else\settowidth{\dimen0}{#1}\fi\hspace*{-\dimen0}}

\newenvironment{center-small}
	{\par\centering\medskip}
	{\par\medskip\noindent}

\usepackage[dvipsnames,hyperref]{xcolor}

\newcommand{\Quad}[1]{
	\mathchoice
	{\quad\text{#1}\quad}
	{\text{ #1 }}
	{\text{ #1 }}
	{\text{ #1 }}
}

\newcommand{\Qand}{\Quad{and}}

\newcommand{\Qfor}{\Quad{for}}
\newcommand{\Qforall}{\Quad{for all}}

\newcommand{\Qif}{\Quad{if}}

\newcommand{\Qwhere}{\Quad{where}}

\NewDocumentCommand{\tv}{sm}{%
	\IfBooleanT{#1}{\bigl}\lVert #2 \IfBooleanT{#1}{\bigr}\rVert_\mathsf{TV}%
}

\newcommand{\one}  [1]{\bm1\{ #1 \}}

\newcommand{\fnrestrict}[2]{\left. {#1} \right|_{#2}}

\newcommand{\bcdot}{\ensuremath{\bm{\cdot}}}
\newcommand{\cq}{\coloneqq}

\newcommand{\abs}  [1]{| #1 |}
\newcommand{\absb} [1]{\bigl| #1 \bigr|}

\newcommand{\rbr} [1]{ ( #1 ) }
\newcommand{\rbb} [1]{\bigl( #1 \bigr)}

\newcommand{\coi}[1]{[#1)}
\newcommand{\oci}[1]{(#1]}

\newcommand{\floor}  [1]{\lfloor #1 \rfloor}

\newcommand{\ceil}[1]  {\lceil #1 \rceil}

\newcommand{\Om}  [1]{\Omega( #1 )}

\newcommand{\Oh}  [1]{\mathcal{O}( #1 )}

\newcommand{\oh}  [1]{o( #1 )}

\newcommand{\Th}  [1]{\Theta( #1 )}
\newcommand{\Thb} [1]{\Theta\bigl( #1 \bigr)}

\providecommand\given{}

\DeclarePairedDelimiterXPP{\PR}[2]
	{\mathbb P_{#1}}{[}{]}{}%
	{\renewcommand\given{\nonscript\:\delimsize\vert\nonscript\:\mathopen{}}#2}
\NewDocumentCommand{\pr}{som}{%
	\IfBooleanTF{#1}
	{\PR[\big]{\IfValueT{#2}{#2}}{#3}}
	{\PR{\IfValueT{#2}{#2}}{#3}}
}

\DeclarePairedDelimiterXPP{\EX}[2]
	{\mathbb E_{#1}}{[}{]}{}%
	{\renewcommand\given{\nonscript\:\delimsize\vert\nonscript\:\mathopen{}}#2}
\NewDocumentCommand{\ex}{som}{%
	\IfBooleanTF{#1}
	{\EX[\big]{\IfValueT{#2}{#2}}{#3}}
	{\EX{\IfValueT{#2}{#2}}{#3}}
}

\DeclarePairedDelimiterXPP{\VAR}[2]
	{\mathbb V\mathrm{ar}_{#1}}{(}{)}{}%
	{\renewcommand\given{\nonscript\:\delimsize\vert\nonscript\:\mathopen{}}#2}
\NewDocumentCommand{\var}{som}{%
	\IfBooleanTF{#1}
	{\VAR[\big]{\IfValueT{#2}{#2}}{#3}}
	{\VAR{\IfValueT{#2}{#2}}{#3}}
}

\let\originalexp\exp
\let\exp\relax
\DeclarePairedDelimiterXPP{\EXP}[1]{\originalexp}{(}{)}{}{#1}
\NewDocumentCommand{\exp}{sm}{%
	\IfBooleanTF{#1}
		{\EXP[\big]{#2}}
		{\EXP{#2}}
}

\DeclarePairedDelimiterX\SET[1]{\{}{\}}%
	{\renewcommand\given{\nonscript\:\delimsize\vert\nonscript\:\mathopen{}}#1}
\NewDocumentCommand{\set}{sm}{%
	\IfBooleanTF{#1}
		{\SET[\big]{#2}}
		{\SET{#2}}
}

\usepackage{calc}
\newlength{\halfplusheight}
\newcommand{\binomt}[2]{
	\mathchoice
	{\textstyle \binom{#1}{#2} \displaystyle}
	{\binom{#1}{#2}}
	{\binom{#1}{#2}}
	{\binom{#1}{#2}}
}

\NewDocumentCommand{\sumt}{smo}{%
	\mathchoice%
	{\textstyle\sum_{#2}\IfBooleanT{#1}{^\star}\IfValueT{#3}{^{#3}}\displaystyle}%
	{\sum_{#2}\IfBooleanT{#1}{^\star}\IfValueT{#3}{^{#3}}}%
	{\sum_{#2}\IfBooleanT{#1}{^\star}\IfValueT{#3}{^{#3}}}%
	{\sum_{#2}\IfBooleanT{#1}{^\star}\IfValueT{#3}{^{#3}}}%
}

\NewDocumentCommand{\sumd}{smo}{%
	\sum_{#2}\IfBooleanTF{#1}{^\star}{\IfValueT{#3}{^{#3}}}%
}

\NewDocumentCommand{\intt}{mo}{%
	\textstyle\int_{#1}\IfNoValueF{#2}{^{#2}}\displaystyle%
}

\NewDocumentCommand{\prodt}{mo}{%
	\mathchoice%
	{\textstyle \prod_{#1}\IfValueT{#2}{^{#2}} \displaystyle}%
	{\prod_{#1}\IfValueT{#2}{^{#2}}}%
	{\prod_{#1}\IfValueT{#2}{^{#2}}}%
	{\prod_{#1}\IfValueT{#2}{^{#2}}}%
}

\NewDocumentCommand{\prodd}{smo}{%
	\prod_{#2}\IfBooleanT{#1}{^\star}\IfValueT{#3}{^{#3}}%
}

\newcommand{\iid}{\textsf{iid}}
\DeclareMathOperator{\Exp}{Exp}

\DeclareMathOperator{\Bern}{Bern}
\DeclareMathOperator{\Bin}{Bin}

\DeclareMathOperator{\Unif}{Unif}

\DeclareMathOperator{\poly}{poly}

\newcommand{\RW}{\ifmmode \mathsf{RW} \else \textsf{RW}\xspace \fi}

\newcommand{\TV}{\ifmmode \mathsf{TV} \else \textsf{TV}\xspace \fi}

\newcommand{\mbn}{\mathbb{N}}

\newcommand{\mbr}{\mathbb{R}}

\newcommand{\mbz}{\mathbb{Z}}

\newcommand{\mcm}{\mathcal{M}}
\newcommand{\mcn}{\mathcal{N}}

\newcommand{\mcq}{\mathcal{Q}}

\newcommand{\mft}{\mathfrak{t}}

\newcommand{\mfC}{\mathfrak{C}}

\newcommand{\eps}{\varepsilon}

\usepackage{manyfoot}
\SetFootnoteHook{\hspace*{-1.8em}}
\DeclareNewFootnote{bl}[gobble]
\newcommand{\blfootnote}[1]{\footnotebl{\sffamily#1}}

\def\IfAmpersandUseAlign#1#2&#3\EndIfAmpersandUseAlign%
	{%
		\if\relax\detokenize{#3}\relax
		\begin{equation*}%
		#1%
		\end{equation*}%
		\else
		\begin{align*}%
		#1%
		\end{align*}%
		\fi
	}
	\def\[#1\]%
	{%
		\IfAmpersandUseAlign{#1}#1&\EndIfAmpersandUseAlign
	}

\usepackage{eqparbox}

\newcommand{\qedtriangle}{\renewcommand{\qedsymbol}{\ensuremath{\triangle}}}

\usepackage[noabbrev, capitalise]{cleveref}

\crefname{figure}{Figure}{Figures}

\numberwithin{equation}{section}

\crefformat{equation}{\textup{(#2#1#3)}}
\crefmultiformat{equation}{\textup{(#2#1#3}}{\textup{, #2#1#3)}}{\textup{, #2#1#3}}{\textup{, #2#1#3)}}
\crefrangeformat{equation}{\textup{(#3#1#4--#5#2#6)}}

\crefformat{section}{\S#2#1#3}
\crefformat{subsection}{\S#2#1#3}
\crefformat{subsubsection}{\S#2#1#3}

\newenvironment{Proof}[1][\proofname]{%
	\proof[\upshape\bfseries\sffamily\boldmath#1]
}{\endproof}

\usepackage{etoolbox}
\usepackage{needspace}

\newtheoremstyle{sfsl}
	{1\baselineskip}		
	{1\baselineskip}		
	{\slshape}				
	{}						
	{\bfseries\sffamily}	
	{.}						
	{0.5em}					
	{\thmname{#1}\thmnumber{ #2}\thmnote{ {\mdseries(#3)}}}

\newtheoremstyle{sfup}
	{1\baselineskip}		
	{1\baselineskip}		
	{\upshape}				
	{}						
	{\bfseries\sffamily}	
	{.}						
	{0.5em}					
	{\thmname{#1}\thmnumber{ #2}\thmnote{ {\mdseries(#3)}}}

\theoremstyle{sfsl}

\newtheorem*{thm*}{Theorem}
\newtheorem{thm} {Theorem}[section]
\crefname{thm}{Theorem}{Theorems}

\newtheorem*{introthm*}{Theorem}

\crefname{introthm}{Theorem}{Theorems}

\newtheorem*{cor*}{Corollary}

\crefname{cor}{Corollary}{Corollaries}

\newtheorem*{introcor*}{Corollary}

\crefname{introcor}{Corollary}{Corollaries}

\newtheorem*{introconj*}{Conjecture}

\crefname{introconj}{Conjecture}{Conjectures}

\newtheorem*{introques*}{Question}

\crefname{introques}{Question}{Questions}

\newtheorem*{lem*}    {Lemma}
\newtheorem{lem} [thm]{Lemma}
\crefname{lem}{Lemma}{Lemmas}

\newtheorem*{introlem*}{Lemma}

\crefname{introlem}{Lemma}{Lemmas}

\newtheorem*{prop*}    {Proposition}
\newtheorem{prop} [thm]{Proposition}
\crefname{prop}{Proposition}{Propositions}

\newtheorem*{openq*}    {Open Question}

\crefname{openq}{Open Question}{Open Questions}

\newtheorem{reduction}[thm]{Reduction}

\newtheorem*{clm*}    {Claim}

\crefname{clm}{Claim}{Claims}

\newtheorem*{defn*}    {Definition}
\newtheorem{defn} [thm]{Definition}
\crefname{defn}{Definition}{Definitions}

\newtheorem*{introdefn*}{Definition}

\crefname{introdefn}{Definition}{Definitions}

\newtheorem*{alg*}{Algorithm}

\crefname{alg}{Algorithm}{Algorithms}

\newtheorem{nota}[thm]{Notation}
\crefname{nota}{Notation}{Notations}

\newtheorem{innercustomgeneric}{\customgenericname}
\providecommand{\customgenericname}{}

\newcommand{\customgenname}{}

\newtheorem*{conj*}   {Conjecture}

\crefname{conj}{Conjecture}{Conjectures}

\newenvironment{conj-ind*}
	{\begin{quote}\textsf{\textbf{Conjecture.}}\slshape}
	{\end{quote}}
\newenvironment{conj-ind}
	{\begin{quote}\vspace{-\glueexpr\baselineskip+\topsep}\begin{customconj}}
	{\end{customconj}\end{quote}}

\newenvironment{question-ind*}
	{\begin{quote}\textsf{\textbf{Question.}}\slshape}
	{\end{quote}}
\newenvironment{question-ind}
	{\begin{quote}\vspace{-\glueexpr\baselineskip+\topsep}\begin{customquestion}}
	{\end{customquestion}\end{quote}}

\newenvironment{openproblem-ind*}
	{\begin{quote}\textsf{\textbf{Open Problem.}}\slshape}
	{\end{quote}}
\newenvironment{openproblem-ind}
	{\begin{quote}\vspace{-\glueexpr\baselineskip+\topsep}\begin{customopenproblem}}
	{\end{customopenproblem}\end{quote}}

\newtheorem*{hypothesis*}{Hypothesis}

\newtheorem*{hyp*}{Hypothesis}

\crefname{hyp}{Hypothesis}{Hypotheses}

\newtheorem*{rmk*}{Remark}

\theoremstyle{sfup}

\crefname{defn} {Definition}{Definitions}
\crefname{defnT}{Definition}{Definitions}

\newtheorem*{defnTT}{Definition}
\newenvironment{defnt*}
	{\pushQED{\qed}\renewcommand{\qedsymbol}{\ensuremath{\triangle}}\defnTT}
	{\popQED\enddefnTT}

\newtheorem{rmkT}[thm]{Remark}
	\newenvironment{rmkt}
	{\pushQED{\qed}\renewcommand{\qedsymbol}{\ensuremath{\triangle}}\rmkT}
	{\popQED\endrmkT}
\crefname{rmk} {Remark}{Remarks}
\crefname{rmkT}{Remark}{Remarks}

\newtheorem*{rmkTT}{Remark}
\newenvironment{rmkt*}
	{\pushQED{\qed}\renewcommand{\qedsymbol}{\ensuremath{\triangle}}\rmkTT}
	{\popQED\endrmkTT}

\crefname{rmks} {Remarks}{Remarks}
\crefname{rmksT}{Remarks}{Remarks}

\newtheorem*{rmks*} {Remarks}
\newtheorem*{rmksTT}{Remarks}
\newenvironment{rmkst*}
	{\pushQED{\qed}\renewcommand{\qedsymbol}{\ensuremath{\triangle}}\rmksTT}
	{\popQED\endrmksTT}

\crefname{intrormk} {Remark}{Remarks}
\crefname{intrormkT}{Remark}{Remarks}

\newtheorem*{intrormk*} {Remark}
\newtheorem*{intrormkTT}{Remark}
\newenvironment{intrormkt*}
	{\pushQED{\qed}\renewcommand{\qedsymbol}{\ensuremath{\triangle}}\intrormkTT}
	{\popQED\endintrormkTT}

\crefname{exm} {Example}{Examples}
\crefname{exmT}{Example}{Examples}

\newtheorem*{exm*} {Example}
\newtheorem*{exmTT}{Lemma}
	\newenvironment{exmt*}
	{\pushQED{\qed}\renewcommand{\qedsymbol}{\ensuremath{\triangle}}\exmTT}
	{\popQED\endexmTT}

\newtheorem*{note*} {Note}
\newtheorem*{noteTT}{Note}
	\newenvironment{notet*}
	{\pushQED{\qed}\renewcommand{\qedsymbol}{\ensuremath{\triangle}}\noteTT}
	{\popQED\endnoteTT}


\newcounter{parentnumber}

\makeatletter
\newenvironment{subtheorem-num}[1]{%
	\def\subtheoremcounter{#1}%
	\refstepcounter{#1}%
	\protected@edef\theparentnumber{\csname the#1\endcsname}%
	\setcounter{parentnumber}{\value{#1}}%
	\setcounter{#1}{0}%
	\expandafter\def\csname the#1\endcsname{\theparentnumber.\arabic{#1}}%
	\expandafter\def\csname theH#1\endcsname{thm.\theparentnumber.\arabic{#1}}%
	\unskip\ignorespaces
}{%
	\setcounter{\subtheoremcounter}{\value{parentnumber}}%
	\ignorespacesafterend
}
\makeatother

\makeatletter
\let\save@mathaccent\mathaccent
\newcommand*\if@single[3]{%
  \setbox0\hbox{${\mathaccent"0362{#1}}^H$}%
  \setbox2\hbox{${\mathaccent"0362{\kern0pt#1}}^H$}%
  \ifdim\ht0=\ht2 #3\else #2\fi
  }
\newcommand*\rel@kern[1]{\kern#1\dimexpr\macc@kerna}
\newcommand*\widebar[1]{\@ifnextchar^{{\wide@bar{#1}{0}}}{\wide@bar{#1}{1}}}
\newcommand*\wide@bar[2]{\if@single{#1}{\wide@bar@{#1}{#2}{1}}{\wide@bar@{#1}{#2}{2}}}
\newcommand*\wide@bar@[3]{%
  \begingroup
  \def\mathaccent##1##2{%
    \let\mathaccent\save@mathaccent
    \if#32 \let\macc@nucleus\first@char \fi
    \setbox\z@\hbox{$\macc@style{\macc@nucleus}_{}$}%
    \setbox\tw@\hbox{$\macc@style{\macc@nucleus}{}_{}$}%
    \dimen@\wd\tw@
    \advance\dimen@-\wd\z@
    \divide\dimen@ 3
    \@tempdima\wd\tw@
    \advance\@tempdima-\scriptspace
    \divide\@tempdima 10
    \advance\dimen@-\@tempdima
    \ifdim\dimen@>\z@ \dimen@0pt\fi
    \rel@kern{0.6}\kern-\dimen@
    \if#31
      \overline{\rel@kern{-0.6}\kern\dimen@\macc@nucleus\rel@kern{0.4}\kern\dimen@}%
      \advance\dimen@0.4\dimexpr\macc@kerna
      \let\final@kern#2%
      \ifdim\dimen@<\z@ \let\final@kern1\fi
      \if\final@kern1 \kern-\dimen@\fi
    \else
      \overline{\rel@kern{-0.6}\kern\dimen@#1}%
    \fi
  }%
  \macc@depth\@ne
  \let\math@bgroup\@empty \let\math@egroup\macc@set@skewchar
  \mathsurround\z@ \frozen@everymath{\mathgroup\macc@group\relax}%
  \macc@set@skewchar\relax
  \let\mathaccentV\macc@nested@a
  \if#31
    \macc@nested@a\relax111{#1}%
  \else
    \def\gobble@till@marker##1\endmarker{}%
    \futurelet\first@char\gobble@till@marker#1\endmarker
    \ifcat\noexpand\first@char A\else
      \def\first@char{}%
    \fi
    \macc@nested@a\relax111{\first@char}%
  \fi
  \endgroup
}
\makeatother

\usepackage{xparse}
\ExplSyntaxOn
\NewDocumentCommand{\mref}{m}{\quinn_mref:n {#1}}
\seq_new:N \l_quinn_mref_seq
\cs_new:Npn \quinn_mref:n #1
{
	\seq_set_split:Nnn \l_quinn_mref_seq { , } { #1 }
	\seq_pop_right:NN \l_quinn_mref_seq \l_tmpa_tl
	( 
	\seq_map_inline:Nn \l_quinn_mref_seq
	{ \ref{##1},\nobreakspace } 
	\exp_args:NV \ref \l_tmpa_tl 
	) 
}
\ExplSyntaxOff

%% file: RandomisedSorting.bib
@incollection{AFHN:adversarial-sorting:conf,
  title = {Sorting and Selection with Imprecise Comparisons},
  booktitle = {Automata, languages and programming. Part I},
  author = {Ajtai, Miklós and Feldman, Vitaly and Hassidim, Avinatan and Nelson, Jelani},
  date = {2009},
  series = {Lecture Notes in Comput. Sci.},
  volume = {5555},
  pages = {37–48},
  publisher = {Springer, Berlin},
  doi = {10.1007/978-3-642-02927-1\_5},
  url = {https://doi.org/10.1007/978-3-642-02927-1_5},
  mrclass = {68P10},
  mrnumber = {2544833}
}

@article{AFHN:adversarial-sorting:jour,
  title = {Sorting and Selection with Imprecise Comparisons},
  author = {Ajtai, Miklós and Feldman, Vitaly and Hassidim, Avinatan and Nelson, Jelani},
  date = {2016},
  journaltitle = {ACM Transactions on Algorithms},
  shortjournal = {ACM Trans. Algorithms},
  volume = {12},
  number = {2},
  pages = {Art. 19, 19},
  issn = {1549-6325,1549-6333},
  doi = {10.1145/2701427},
  url = {https://doi.org/10.1145/2701427},
  fjournal = {ACM Transactions on Algorithms},
  mrclass = {68P10 (68Q17 68Q25 68W05)},
  mrnumber = {3465942},
  mrreviewer = {MohammadFarshi}
}

@article{AHRV:random-sorting-networks,
  title = {Random Sorting Networks},
  author = {Angel, Omer and Holroyd, Alexander E. and Romik, Dan and Virág, Bálint},
  date = {2007},
  journaltitle = {Advances in Mathematics},
  shortjournal = {Adv. Math.},
  volume = {215},
  number = {2},
  pages = {839–868},
  issn = {0001-8708,1090-2082},
  doi = {10.1016/j.aim.2007.05.019},
  url = {https://doi.org/10.1016/j.aim.2007.05.019},
  fjournal = {Advances in Mathematics},
  mrclass = {60C05 (05C25 05E10 68P10)},
  mrnumber = {2355610},
  mrreviewer = {YoshiharuKohayakawa}
}

@article{AHS:counting-networks,
  title = {Counting Networks},
  author = {Aspnes, James and Herlihy, Maurice and Shavit, Nir},
  date = {1994-09},
  journaltitle = {Journal of The Acm},
  shortjournal = {J. ACM},
  volume = {41},
  number = {5},
  pages = {1020–1048},
  publisher = {Association for Computing Machinery},
  location = {New York, NY, USA},
  issn = {0004-5411},
  doi = {10.1145/185675.185815},
  url = {https://doi.org/10.1145/185675.185815},
  issue_date = {Sept. 1994},
  pagetotal = {29}
}

@inproceedings{AKS:sorting:conf,
  title = {An $O(n \log n)$ Sorting Network},
  booktitle = {Proceedings of the Fifteenth Annual ACM Symposium on Theory of Computing},
  author = {Ajtai, M. and Komlós, J. and Szemerédi, E.},
  date = {1983},
  series = {STOC '83},
  pages = {1–9},
  publisher = {Association for Computing Machinery},
  location = {New York, NY, USA},
  doi = {10.1145/800061.808726},
  url = {https://doi.org/10.1145/800061.808726},
  isbn = {0-89791-099-0},
  pagetotal = {9}
}

@article{AKS:sorting:jour,
  title = {Sorting in $c \log n$ Parallel Steps},
  author = {Ajtai, M. and Komlós, J. and Szemerédi, E.},
  date = {1983},
  journaltitle = {Combinatorica},
  shortjournal = {Combinatorica},
  volume = {3},
  number = {1},
  pages = {1–19},
  issn = {0209-9683},
  doi = {10.1007/BF02579338},
  url = {https://doi.org/10.1007/BF02579338},
  fjournal = {Combinatorica},
  mrclass = {68P10},
  mrnumber = {716418}
}

@inproceedings{B:bitonic,
  title = {Sorting Networks and Their Applications},
  booktitle = {Proceedings of the April 30–May 2, 1968, Spring Joint Computer Conference},
  author = {Batcher, K. E.},
  date = {1968},
  series = {AFIPS '68 (Spring)},
  pages = {307–314},
  publisher = {Association for Computing Machinery},
  location = {New York, NY, USA},
  doi = {10.1145/1468075.1468121},
  url = {https://doi.org/10.1145/1468075.1468121},
  isbn = {978-1-4503-7897-0},
  pagetotal = {8}
}

@article{C:parallel-mergesort,
  title = {Parallel Merge Sort},
  author = {Cole, Richard},
  date = {1988},
  journaltitle = {SIAM Journal on Computing},
  shortjournal = {SIAM J. Comput.},
  volume = {17},
  number = {4},
  pages = {770–785},
  issn = {0097-5397},
  doi = {10.1137/0217049},
  url = {https://doi.org/10.1137/0217049},
  fjournal = {SIAM Journal on Computing},
  mrclass = {68P10 (68Q10)},
  mrnumber = {953293},
  mrreviewer = {VictorS.Grinberg}
}

@inproceedings{GSS:prob-bubble-sort,
  title = {New Clocks, Optimal Line Formation and Self-Replication Population Protocols},
  booktitle = {40th International Symposium on Theoretical Aspects of Computer Science (STACS 2023)},
  author = {Gąsieniec, Leszek and Spirakis, Paul G. and Stachowiak, Grzegorz},
  editor = {Berenbrink, Petra and Bouyer, Patricia and Dawar, Anuj and Kanté, Mamadou Moustapha},
  date = {2023},
  series = {Leibniz International Proceedings in Informatics (LIPIcs)},
  volume = {254},
  pages = {33:1–33:22},
  publisher = {Schloss Dagstuhl – Leibniz-Zentrum für Informatik},
  location = {Dagstuhl, Germany},
  issn = {1868-8969},
  doi = {10.4230/LIPIcs.STACS.2023.33},
  url = {https://drops.dagstuhl.de/entities/document/10.4230/LIPIcs.STACS.2023.33},
  isbn = {978-3-95977-266-2},
  mrnumber = {4587217},
  urn = {urn:nbn:de:0030-drops-176857}
}

@article{I:random-sorting-network-se,
  title = {Probability That a Random Sorting Network Works},
  author = {Irving, Geoffrey},
  date = {2018-02-28},
  journaltitle = {Theoretical Computer Science Stack Exchange},
  url = {https://cstheory.stackexchange.com/q/40293},
  customeprint = {https://cstheory.stackexchange.com/q/40293},
  howpublished = {Theoretical Computer Science Stack Exchange}
}

@article{M:uniform-sort,
  title = {Answer to ``What is the expected number of comparisons of a sorting algorithm that chooses a random pair of elements and swaps if out of order, until sorted?''},
  author = {Murty, Sanjeev},
  date = {2019},
  journaltitle = {Quora},
  url = {https://www.quora.com/What-is-the-expected-number-of-comparisons-of-a-sorting-algorithm-that-chooses-a-random-pair-of-elements-and-swaps-if-out-of-order-until-sorted/answer/Sanjeev-Murty},
  urldate = {2024-12-11},
  customeprint = {https://qr.ae/pYR8yl},
  langid = {english}
}

@book{S:algorithm-design-manual,
  title = {The Algorithm Design Manual},
  author = {Skiena, Steven S.},
  date = {2020},
  series = {Texts in Computer Science},
  edition = {3},
  pages = {viii+793},
  publisher = {Springer, Cham},
  doi = {10.1007/978-3-030-54256-6},
  url = {https://doi.org/10.1007/978-3-030-54256-6},
  mrclass = {68-01 (68P05 68Wxx)},
  mrnumber = {4241430}
}

@inproceedings{T:sorting-selection-adversary,
  title = {Sorting and Selection in Rounds with Adversarial Comparisons},
  booktitle = {Proceedings of the 2024 Annual ACM-SIAM Symposium on Discrete Algorithms (SODA)},
  author = {Trevisan, Christopher},
  date = {2024},
  pages = {1099–1119},
  publisher = {SIAM, Philadelphia, PA},
  doi = {10.1137/1.9781611977912.42},
  url = {https://doi.org/10.1137/1.9781611977912.42},
  isbn = {978-1-61197-791-2},
  mrclass = {68P10},
  mrnumber = {4699287}
}
